\documentclass{article}
\usepackage[T1]{fontenc}
\usepackage{graphics}

\makeatletter

\providecommand{\LyX}{L\kern-.1667em\lower.25em\hbox{Y}\kern-.125emX\@}

\usepackage[T1]{fontenc}
\usepackage{bookman}
\usepackage{geometry}
\usepackage{graphics}

\makeatletter

\usepackage{a4}
\input{epsfig.sty}
\def\fnum@table{\tablename~{\bf\thetable}}
\def\fnum@figure{\figurename~{\bf\thefigure}}
\def\tablename{\footnotesize{\bf Table}}
\def\figurename{\footnotesize{\bf Figure}}

\def\be{\begin{equation}}
\def\ee{\end{equation}}
\makeatother

\makeatother

\begin{document}

\title{\textbf{\Huge Physics of Event Generators} \thanks{
Invited lecture, given at the Pan-American Advanced Study Institute ``New States
of Matter in Hadronic Interactions'', Campos de Jordao, Brazil, January 7-18,
2002
}}

\author{\textbf{K. Werner}}

\date{\vspace*{-1cm}\protect\( \qquad \protect \)}

\maketitle
{\par\centering \textit{\small SUBATECH, Université de Nantes -- IN2P3/CNRS --
EMN,  Nantes, France }\small \par}
\vspace{0.7cm}

\begin{abstract}
An event generator for nuclear collisions is a microscopic model, obtained from
extrapolating elementary interactions -- as electron-positron annihilation,
deep inelastic scattering, and proton-proton interactions -- towards proton-nucleus
and nucleus-nucleus scattering, by using Monte Carlo techniques. 

In this paper, we will discuss the physical concepts behind such event generators.
We first present some qualitative discussion of nuclear scattering, before discussing
particle production and strings. We then discuss the parton model, and finally
multiple scattering theory. 
\end{abstract}

\section{Qualitative Discussion of Nuclear Scattering}

\subsection{Overview}

\begin{figure}[htb]
{\par\centering \resizebox*{!}{0.15\textheight}{\includegraphics{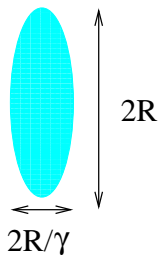}} \par}

\caption{Nuclei are Lorentz contracted. \label{reac1}}
\end{figure}

Relativistic nuclei are Lorentz contracted, which means that the longitudinal
dimension \( 2R \) is reduced to \( 2R/\gamma  \), see fig. \ref{reac1}, 
where \( R \) is the nuclear radius, and \( \gamma =1/\sqrt{1-(v/c)^{2}} \)
the so-called gamma factor. At the heavy ion collider RHIC, we have \( \gamma =100 \)
and at LHC about \( \gamma =3000 \), so relativistic contraction plays a very
essential role.  

\begin{figure}[htb]
{\par\centering \resizebox*{!}{0.15\textheight}{\includegraphics{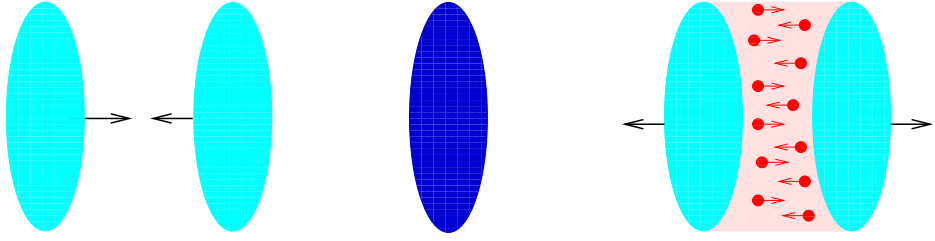}} \par}

\caption{Nuclei pass through each other in a very short time. \label{reac4}}
\end{figure}

Considering the collision of two nuclei, there are first of all the primary
interactions, when the two nuclei pass trough each other in a very short time,
see fig \ref{reac4}.
Since at very high energies the longitudinal size is due to the gamma factor
almost zero (of the order of 0.1 fm at RHIC and 0.01 fm at LHC), all the nucleons
of the projectile interact with all the nucleons of the target instantaneously.
Many elementary interactions between nucleons in the two nuclei happen in parallel,
resulting in many partons (quarks and gluons), moving mainly in longitudinal
direction (pre-equilibrium). These partons interact
\begin{figure}[htb]
{\par\centering \resizebox*{!}{0.15\textheight}{\includegraphics{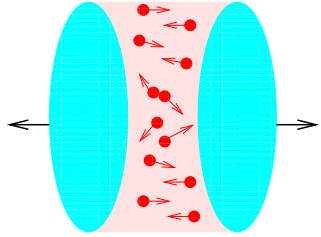}} \par}

\caption{The partons interact and finally reach equilibrium (QGP). \label{reac5}}
\end{figure}
 and finally reach equilibrium, referred to as quark-gluon plasma (QGP). The
system then expands, passing via a phase transition (or sudden crossover) into
the hadronic phase. The density decreases further till the collision rate is
no longer large enough to maintain chemical equilibrium, but there are still
hadronic interactions till finally the particles ``freeze out'', i.e. they
continue their way without further interactions.

Unfortunately there does not exist a single formalism being able to account
for a complete nucleus-nucleus collision. Rather we have - at least for the
moment - to divide the reaction into different stages, and try to understand
the different stages as good as possible. These different stages are 

\begin{itemize}
\item Initial stage, 
\item Pre-equilibrium stage, 
\item Quark-gluon plasma, 
\item Phase transition, 
\item Hadron gas, 
\item Non-equilibrium hadronic matter, 
\item Free hadrons. 
\end{itemize}
Before discussing these stages one after the other, we introduce some useful
variables, as there are the proper time \( \tau  \) and the space time rapidity
\( \eta  \) 
\[
\tau =\sqrt{t^{2}-z^{2}},\qquad \eta =\frac{1}{2}\ln \frac{t+z}{t-z},\]
and the transverse mass \( m_{t} \) and the rapidity \( y \)
\[
m_{t}=\sqrt{E^{2}-p_{z}^{2}},\qquad y=\frac{1}{2}\ln \frac{E+p_{z}}{E-p_{z}}.\]
 The proper time and the transverse mass have the property to be invariant under
Lorentz transformations. The (space time) rapidity is additive under Lorentz
boosts. 

\begin{figure}[htb]
{\par\centering \resizebox*{!}{0.2\textheight}{\includegraphics{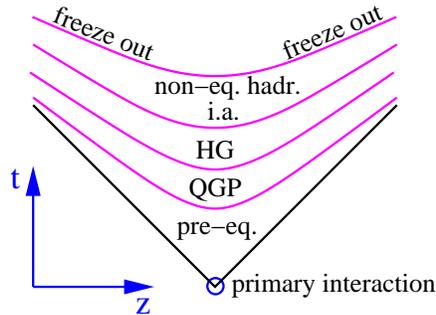}} \par}

\caption{The different stages of heavy ion collisions.\label{stages}}
\end{figure}

Knowing that constant proper time represents hyperbolas in space-time (\( t-z) \),
one may present a space-time picture of the different stages of heavy ion collisions,
as shown in fig. \ref{stages}.

\subsection{Initial Stage}

\begin{figure}[htb]
{\par\centering \resizebox*{!}{0.15\textheight}{\includegraphics{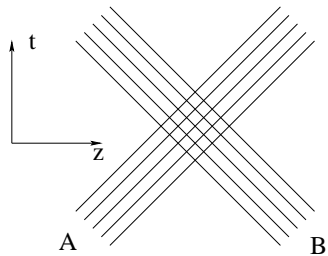}} \par}

\caption{The initial stage of heavy ion collisions.         \label{ini-sta}}
\end{figure}

The understanding of the initial interactions is crucial for any theoretical
treatment of a possible parton-hadron phase transition, the detection of which
being the ultimate aim of all the efforts of colliding heavy ions at very high
energies. Theoretical approaches have to consider the fact that the nuclear
collision happens on a very short time scale, such that all nucleons of the
target interact with all nucleons of the projectile practically instantaneous,
see fig. \ref{ini-sta}. 

It is quite clear that coherence is crucial for the very early stage of nuclear
collisions, so a real quantum treatment is necessary and any attempt to use
a transport theoretical parton approach with incoherent quasi-classical partons
should not be considered at this point. Also semi-classical hadronic cascades
cannot be stretched to account for the very first interactions, even when this
is considered to amount to a string excitation, since it is well known \cite{dre00}
that such a longitudinal excitation is simple kinematically impossible.

So what are the currently used fully quantum mechanical approaches? There are
presently considerable efforts to describe nuclear collisions via solving classical
Yang-Mills equations, which allows to calculate inclusive parton distributions
\cite{mcl94}. This approach is to some extent orthogonal to ours: here, screening
is due to perturbative processes, whereas we claim to have good reasons to consider
soft processes to be at the origin of screening corrections.

Provided factorization works for nuclear collisions, on may employ the parton
model, which allows to calculate inclusive cross sections as a convolution of
an elementary cross section with parton distribution functions, with these distribution
functions taken from deep inelastic scattering. Parton model based are for example
Pythia \cite{sjo87} and HIJING \cite{wan96}.

Another approach is the so-called Gribov-Regge theory. This is an effective
field theory, which allows multiple interactions to happen ``in parallel'',
with the phenomenological object called ``Pomeron'' representing an elementary
interaction. Using the general rules of field theory, on may express cross sections
in terms of a couple of parameters characterizing the Pomeron. A disadvantage
is the fact that cross sections and particle production are not calculated consistently:
the fact that energy needs to be shared between many Pomerons in case of multiple
scattering is well taken into account when calculating particle production (Monte
Carlo applications), but energy conservation is not taken care of for cross
section calculations. Models based on this approach are QGS \cite{kai82}, DPM
\cite{cap94,aur94}, and VENUS \cite{wer93}.

A new approach, called ``Parton-based Gribov-Regge Theory'' \cite{dre00},
solves some of the above-mentioned problems: one has a consistent treatment
for calculating cross sections and particle production, considering energy conservation
in both cases; one introduces hard processes in a natural way, and, compared
to the parton model, one can deal with total cross sections without arbitrary
assumptions. This model is incorporated in the NEXUS \cite{dre00} event generator.

\subsection{Pre-equilibrium Stage}

The partons created in the primary interactions are certainly far from equilibrium,
and is desirable to understand microscopically the equilibrium of the system,
in other words the formation of a quark gluon plasma. This is a difficult task,
since for example at RHIC energies there is still a large soft component. Nevertheless
it is useful to study the evolution of partonic systems based on pQCD, ignoring
soft physics.

\begin{figure}[htb]
{\par\centering \resizebox*{!}{0.15\textheight}{\includegraphics{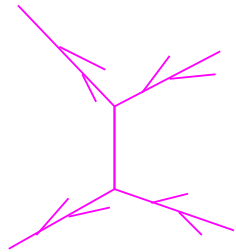}} \par}

\caption{Partons, which have been produced initially, interact. \label{pc}}
\end{figure}

The theoretical tool for this stage is the ``parton cascade'', which amounts
to considering partons as classical particles which move on straight line trajectories,
where binary interactions are defined via parton-parton cross sections calculated
in the framework of perturbative QCD \cite{gei92}, see fig. \ref{pc}. 

One has to carefully regard the range of validity of this approach: it is not
meant to treat the primary interactions, where quantum mechanical interference
should play a crucial role, so one may start the calculation once a system of
incoherent classical partons has been established. On the other end, one should
not stretch the perturbative treatment too far: perturbative calculations require
large momentum transfer, which is not any more guaranteed if the interaction
energy is getting too low.

\subsection{Equilibrium Stage }

We are now discussing the final stage of the collision, consisting of the QGP
phase, the phase transition, and the hadron gas phase. We do not treat these
three stages individually, because the known models treat usually more than
just one stage.

The final aim of all the efforts in the field of ultra-relativistic heavy ion
collisions is the creation of a thermalized system of quarks and gluons. Provided
such an equilibrium has been established, one may use hydrodynamics, which is
a macroscopic approach based on energy-momentum conservation and local thermal
equilibrium. Hydrodynamical calculations have been used since a long time, either
assuming particular symmetries and using analytical methods \cite{bay84}, or
full 3-dimensional calculations numerical calculations \cite{ris98}. Recently
a new technique has been proposed, the so-called smoothed particle hydrodynamics
\cite{agu00}, where fields \( \rho (x) \) are represented by particles as
\( \rho _{P}(x)=\Sigma _{b}\nu _{b}\delta (x-x_{b}) \), and then smoothed:

\[
\rho (x)\rightarrow \rho _{SP}(x)=\int \rho _{P}(x)W(x-x')dx'=\Sigma _{b}\nu _{b}W(x-x_{b})\]
 with some smoothing kernel W. The advantage is that the hydrodynamical equations
are transfered into a system of ordinary differential equations, which can be
solved by applying standard methods. In this way one may perform 3-dimensional
calculations much faster than with traditional methods.

There are several attempts to treat at least the region around the phase transition
in a microscopic way. A possibility is to apply transport theory based on the
NJL model \cite{reh98}, which is an effective theory with a point-like interaction
between two quarks (gluons are not considered explicitly). The model allows
also for hadron production like quark plus anti-quark goes into meson plus meson.
The dynamics is crucially affected by the density and temperature dependence
of quark and hadron masses, one observes for example the formation of droplets
of quark matter rather than homogeneous matter of lower density, since the latter
one would imply higher quark masses.

A completely different hadronization scenario has been proposed based on the
confinement mechanism \cite{hof99}, again ignoring gluons. Quarks are considered
to be classical particles, their dynamics being determined by a classical Hamiltonian.
The latter one contains a string potential and color factors, which force the
quarks to form resonances, which subsequently decay into hadrons.

Another alternative approach is the hadronization via coalescence \cite{csi99}.
Again, starting from a quark-anti-quark plasma, hadronic resonances are formed
based on coalescence, with a subsequent decay into hadrons.

\subsection{Post-equilibrium Hadronic Stage}

Once a purely hadronic system has been established, a microscopic treatment
based on binary hadronic interactions is feasible. Here, hadrons propagate on
classical trajectories and interact according to hadron-hadron scattering cross
sections. If possible, parameterizations of measured cross sections are used.
A couple of models have been constructed along these lines, like UrQMD \cite{bas98,ble99},
ART \cite{li95}, JAM \cite{nar97}. Unfortunately, not all the necessary cross
sections have been measured to a sufficient precision, and correspondingly,
the above-mentioned approaches differ by using different model assumptions for
the cross sections. We emphasize again that hadronic transport codes are a useful
tool to treat the final stage of a heavy ion collision, but not for the primary
interaction.

\section{Particle Production, Strings}

Particle production is relevant for \( pp \), \( pA \) and the initial stage
of \( AA \) collisions. However it is necessary to first study simpler systems
as electron-positron annihilation.

\subsection{The String Picture}

\begin{figure}[htb]
{\par\centering \resizebox*{!}{0.12\textheight}{\includegraphics{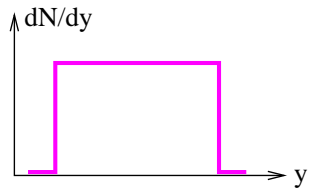}} \par}

\caption{One observes (idealized) a rapidity plateau.         \label{ee1}}
\end{figure}

\begin{figure}[htb]
{\par\centering \resizebox*{!}{0.12\textheight}{\includegraphics{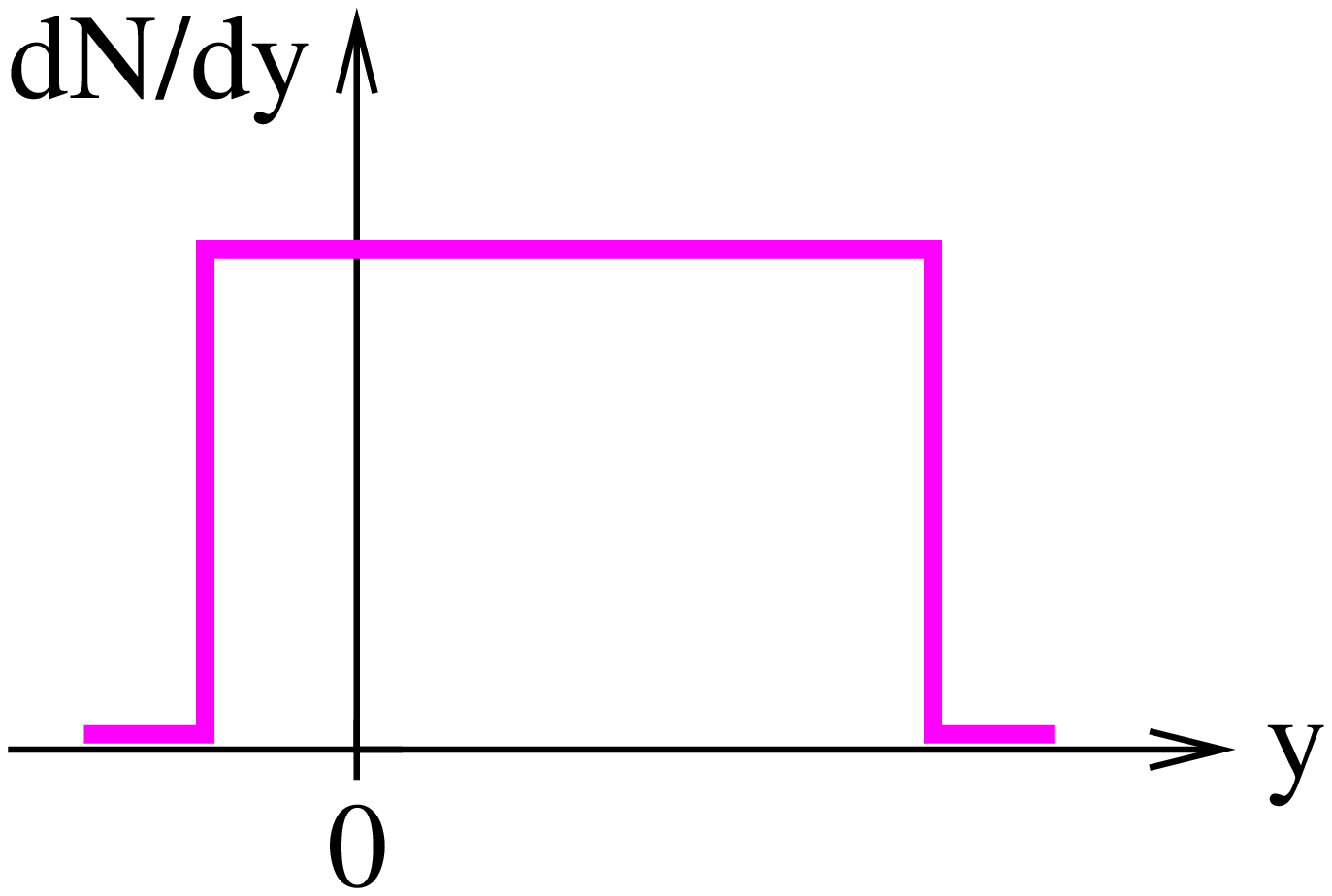}} \resizebox*{!}{0.12\textheight}{\includegraphics{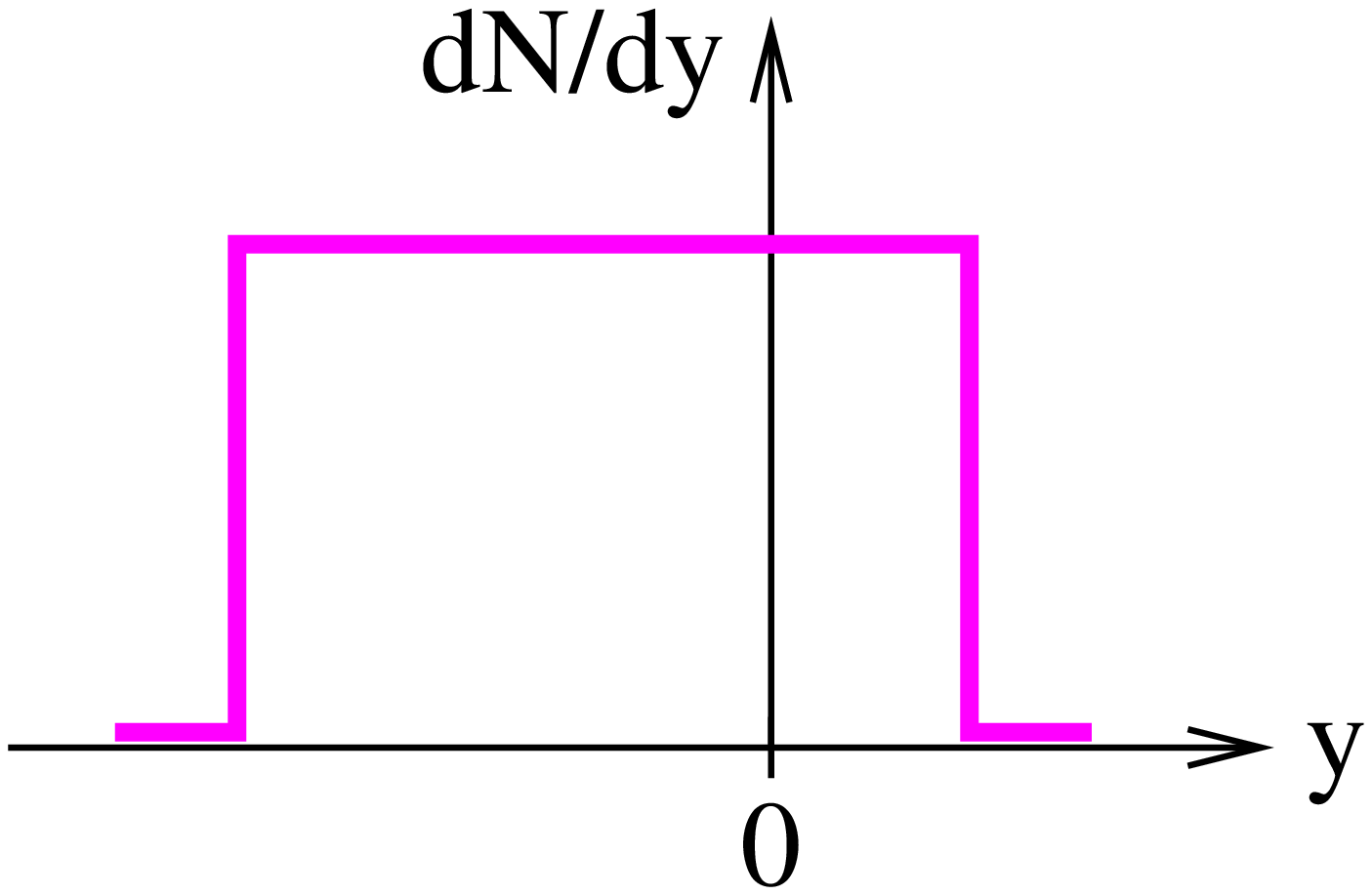}} \par}

\caption{Manifestation of boost invariance         \label{ee3}}
\end{figure}

\begin{figure}[htb]
{\par\centering \resizebox*{!}{0.04\textheight}{\includegraphics{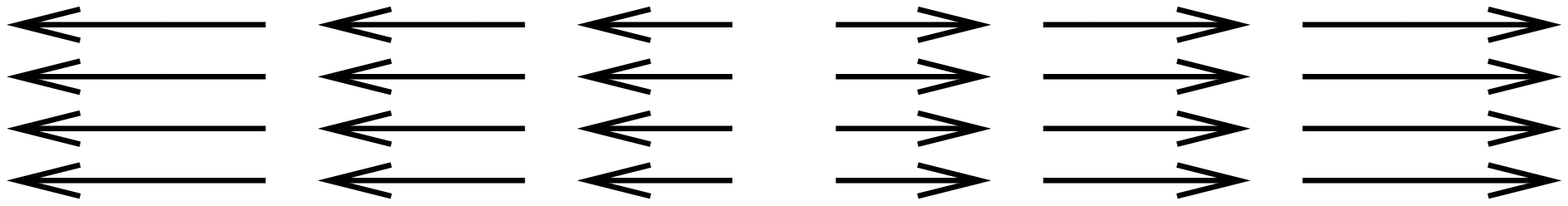}} \( \qquad  \)
\resizebox*{!}{0.04\textheight}{\includegraphics{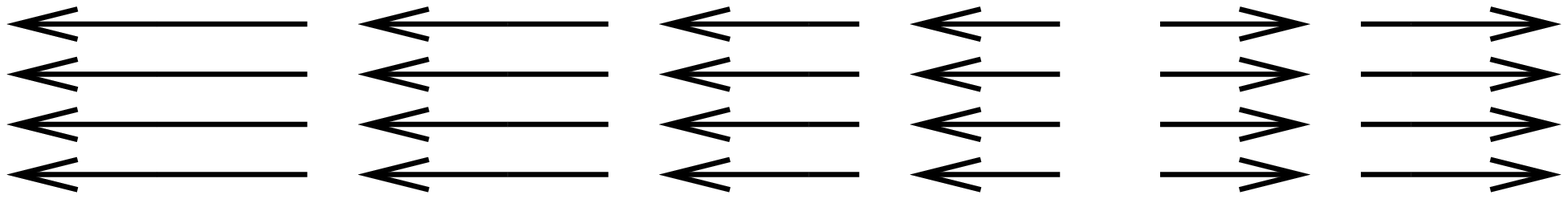}} \par}

\caption{Manifestation of boost invariance         \label{ee5}}
\end{figure}

For \( e^{+}e^{-} \) collisions, we have data available in a wide energy range
(up to 200 GeV in the cms). Studying particle production, one observes (idealized)
a rapidity plateau, see fig. \ref{ee1}. 
When we move the reference system (for example from lab to rest frame), we observe
a manifestation of boost invariance: the same rapidity distribution before and
after the boost, see fig \ref{ee3}. 
What does boost invariance mean? Suppose an expanding dynamical system such
that some central part is at rest and the outer parts move away from the center,
with increasing speed for larger distances, see fig. \ref{ee5}(left). 
 Now we perform a boost such that a different piece of the system is at rest.
In the neighborhood of this region the system looks identical to the neighborhood
of the point at rest before the boost, see fig. \ref{ee5}(right). In other
words: the system is identical at all points in the corresponding local comoving
frame. 

\begin{figure}[htb]
{\par\centering (a)\( \,  \)\resizebox*{!}{0.12\textheight}{\rotatebox{90}{\includegraphics{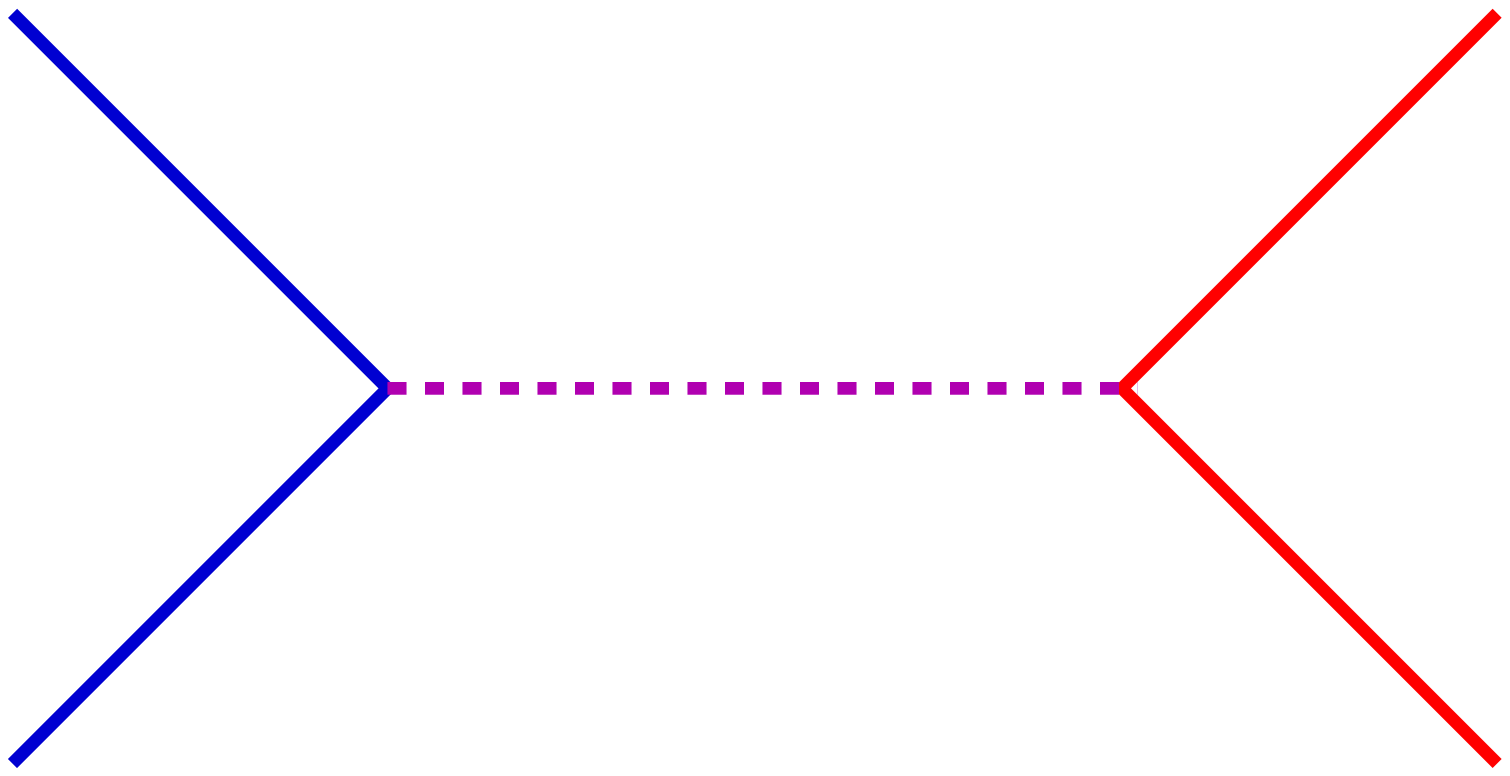}}} \( \qquad  \)(b)\resizebox*{!}{0.03\textheight}{\includegraphics{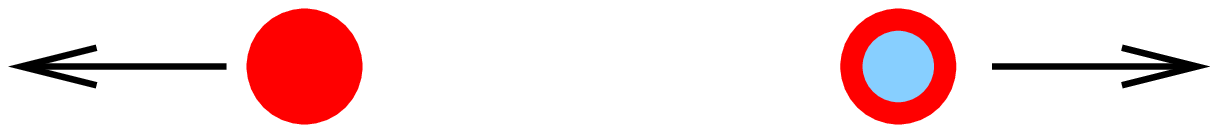}} \par}

\caption{(a) Electron and positron annihilate and form a virtual photon, then the virtual
photon decays into a quark-antiquark pair. (b) Quark and antiquark move apart
from each other.\label{ee8}}
\end{figure}

\begin{figure}[htb]
{\par\centering \resizebox*{!}{0.03\textheight}{\includegraphics{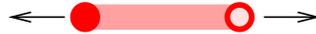}}  \par}

\caption{The quark-antiquark pair forms a string.         \label{ee7}}
\end{figure}

\begin{figure}[htb]
{\par\centering \resizebox*{!}{0.03\textheight}{\includegraphics{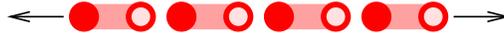}} \par}

\caption{The string breaks via \char`\"{}quark-antiquark\char`\"{} production.         \label{ee9}}
\end{figure}

\begin{figure}[htb]
{\par\centering before:\resizebox*{!}{0.012\textheight}{\includegraphics{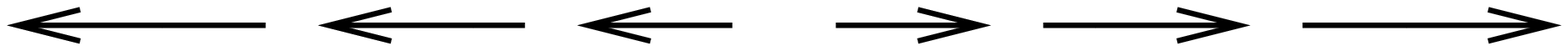}} \( \:  \)after:\resizebox*{!}{0.012\textheight}{\includegraphics{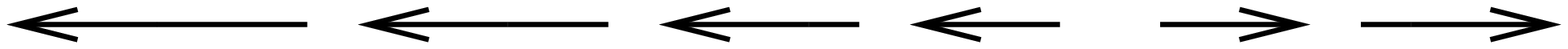}} \par}

\caption{String fragmentation is a boost invariant procedure. \label{ee10b}}
\end{figure}

\begin{figure}[htb]
{\par\centering \resizebox*{!}{0.2\textheight}{\includegraphics{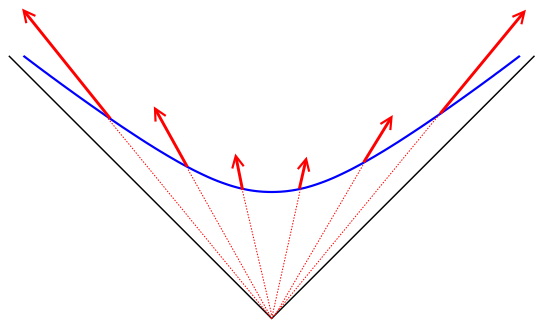}} \par}

\caption{Space-time picture of string decay.     \label{ee12}}
\end{figure}

What happens really? Electron and positron annihilate and form a virtual photon,
then the virtual photon decays into a quark-antiquark pair, see fig. \ref{ee8}(a).
The quark and antiquark move apart from each other, see fig. \ref{ee8}(b).
But quarks and antiquarks cannot be observed individually! There is a gluon
field acting between the two, whose energy is proportional to the separation
distance. This object is called string, see fig. \ref{ee7}.
To separate the quark from the antiquark, one need an infinite energy, which
is impossible. The string breaks via quark-antiquark production, and these new
string pieces are finally hadrons or resonances, see fig. \ref{ee9}. 
String fragmentation is a boost invariant procedure and provides exactly the
situation discussed above: seen from a given point on the string, all the string
pieces move away from this point with increasing speed towards the edges, as
indicated in fig. \ref{ee10b}. In fig. \ref{ee12}, 
we present the space-time picture of the string dynamics: at given proper time
(hyperbola), the velocities of the string pieces (arrows) are such that they
point all back to the origin and are longer towards the edges. This string decay
provides a flat rapidity distribution.

\subsection{What is Really Done}

\begin{figure}[htb]
{\par\centering \resizebox*{!}{0.3\textheight}{\includegraphics{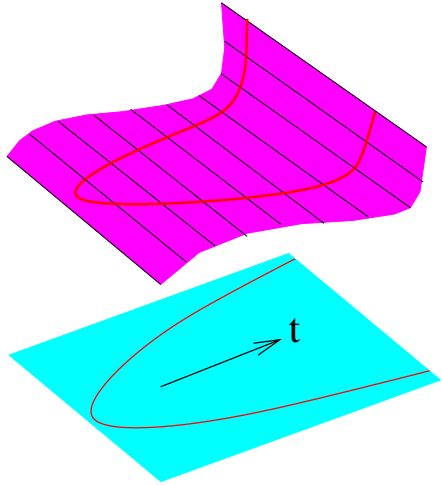}} \par}

\caption{The string surface.       \label{stringtau}}
\end{figure}

\begin{figure}[htb]
{\par\centering \resizebox*{!}{0.4\textheight}{\includegraphics{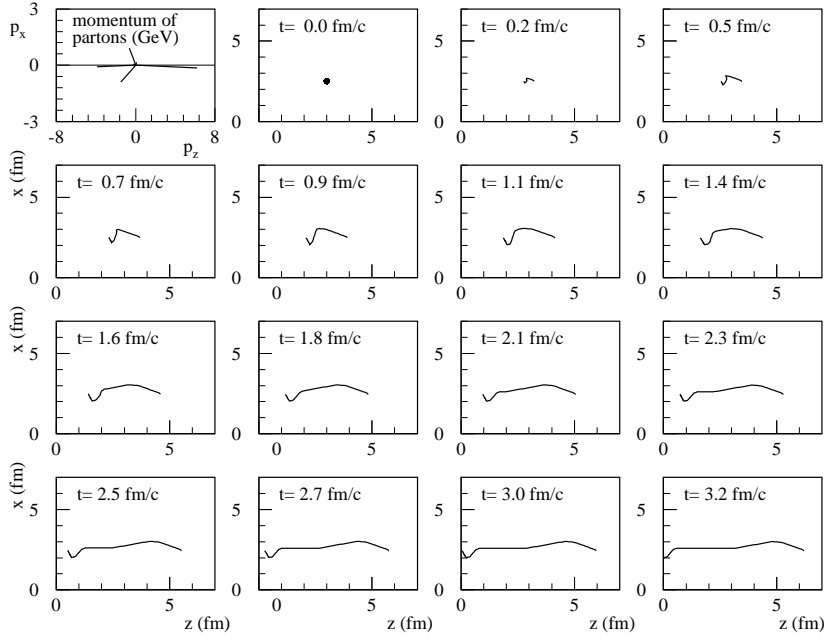}} \par}

\caption{String evolution.         \label{kstring}}
\end{figure}

After this qualitative discussion, let us discuss what is really done. A string
can be considered as a two-dimensional surface in Minkowski space 

\[
X=X(r,t),\]
 with \( r \) being a space-like and \( t \) a time-like parameter, see fig.
\ref{stringtau}.
In order to obtain the equations of motion, we need a Lagrangian. It is obtained
by demanding the invariance of the action with respect to gauge transformations.
This way one finds \cite{dre00} the Lagrangian of Nambu-Goto:

\[
L=-\kappa \sqrt{(X'\dot{X})^{2}-X'^{2}{\dot{X}}^{2},}\]
 with ``dot'' and ``prime'' referring to the partial derivatives with respect
to \( r \) and \( t \), and with \( \kappa  \) being the string tension.
With this Lagrangian we get the Euler-Lagrange equations of motion:

\[
\frac{\partial }{\partial t}\frac{\partial L}{\partial \dot{X}_{\mu }}+\frac{\partial }{\partial r}\frac{\partial L}{\partial X'_{\mu }}=0.\]
We use the gauge fixing

\[
X'^{2}+{\dot{X}}^{2}=0\: \mathrm{and}\: X'\dot{X}=0,\]
which provides a very simple equation of motion, namely a wave equation,

\[
\frac{\partial ^{2}X_{\mu }}{\partial t^{2}}-\frac{\partial ^{2}X_{\mu }}{\partial r^{2}}=0,\]
 with the boundary conditions: \( \partial X_{\mu }/\partial \sigma =0,\: \sigma =0,\pi  \).
The solution of the equation of motion (with initial extension zero) is

\[
X^{\mu }(r,t)=X_{0}+\frac{1}{2}\left( \int _{r-t}^{r+t}g^{\mu }(\xi )d\xi \right) ,\]
 where \( g \) is the initial velocity, \( g(r)=\dot{X}(r,t)_{t=0} \) . Strings
are classified according to the function \( g \). Strings with piecewise constant
\( g \) are called kinky strings, each segment being called kink, finally identified
with perturbative partons. In fig. \ref{kstring}, we show the evolution of
a string generated in electron-positron annihilation (3 internal kinks). 

\vspace{1cm}
\subsection{Results}

\begin{figure}[htb]
{\par\centering \resizebox*{!}{0.24\textheight}{\includegraphics{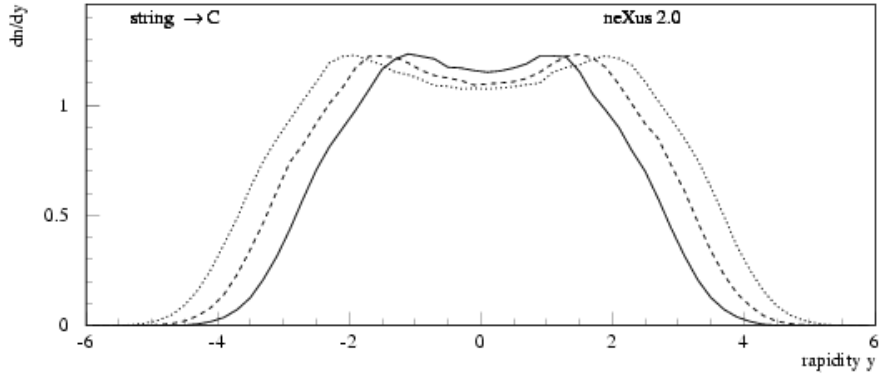}} \par}

\caption{Rapidity distributions of flat strings at 14-22-34 GeV.         \label{ree3}}
\end{figure}

We show some results for rapidity distributions from flat strings (no internal
kinks) in fig. \ref{ree3}. We observe a nice rapidity plateau, which gets broader
with increasing energy. But the plateau height stays constant. Increasing the
energy does not change the local properties of the string, the number of particles
per unit of rapidity stays constant.

\begin{figure}[htb]
{\par\centering \resizebox*{!}{0.42\textheight}{\includegraphics{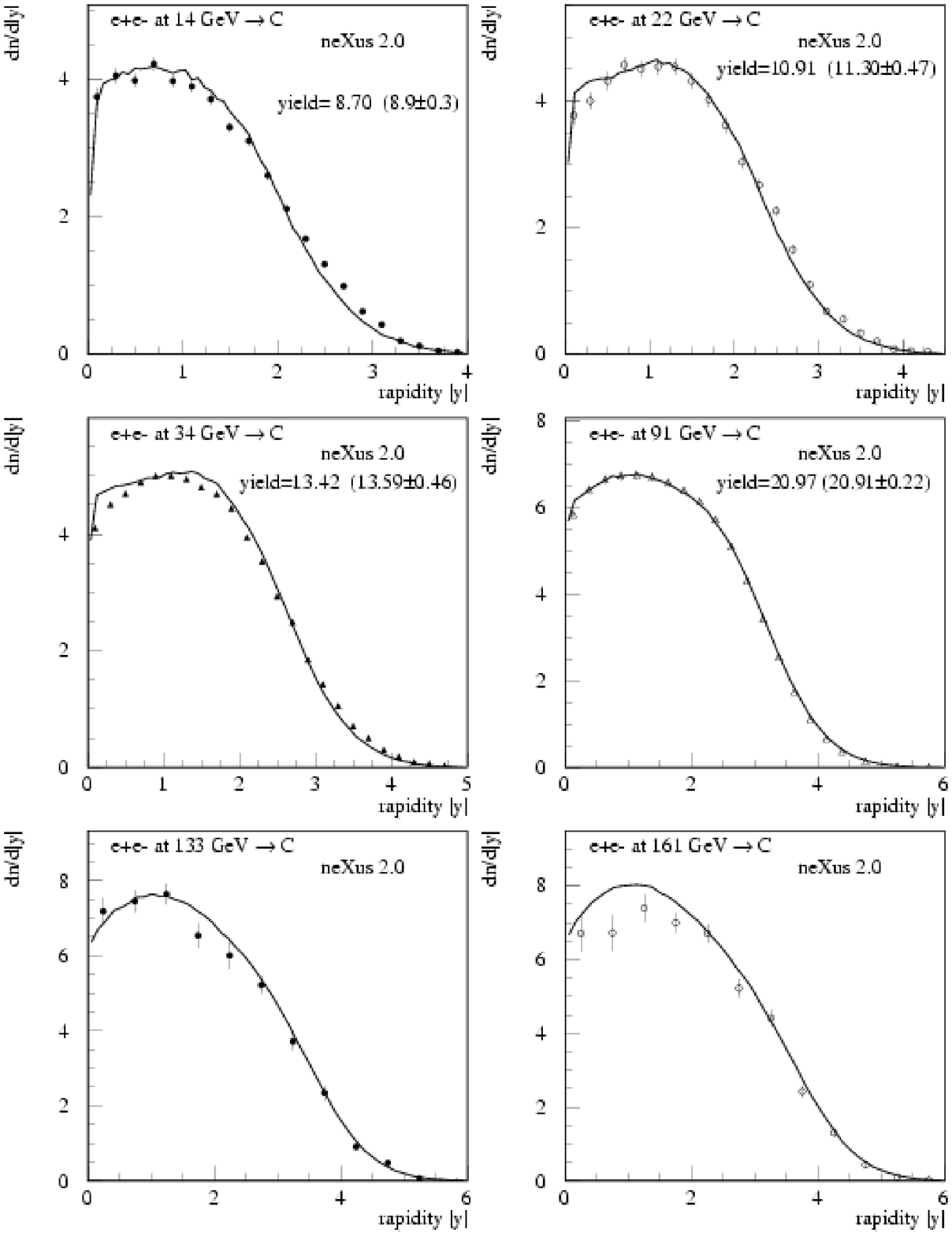}} \resizebox*{!}{0.42\textheight}{\includegraphics{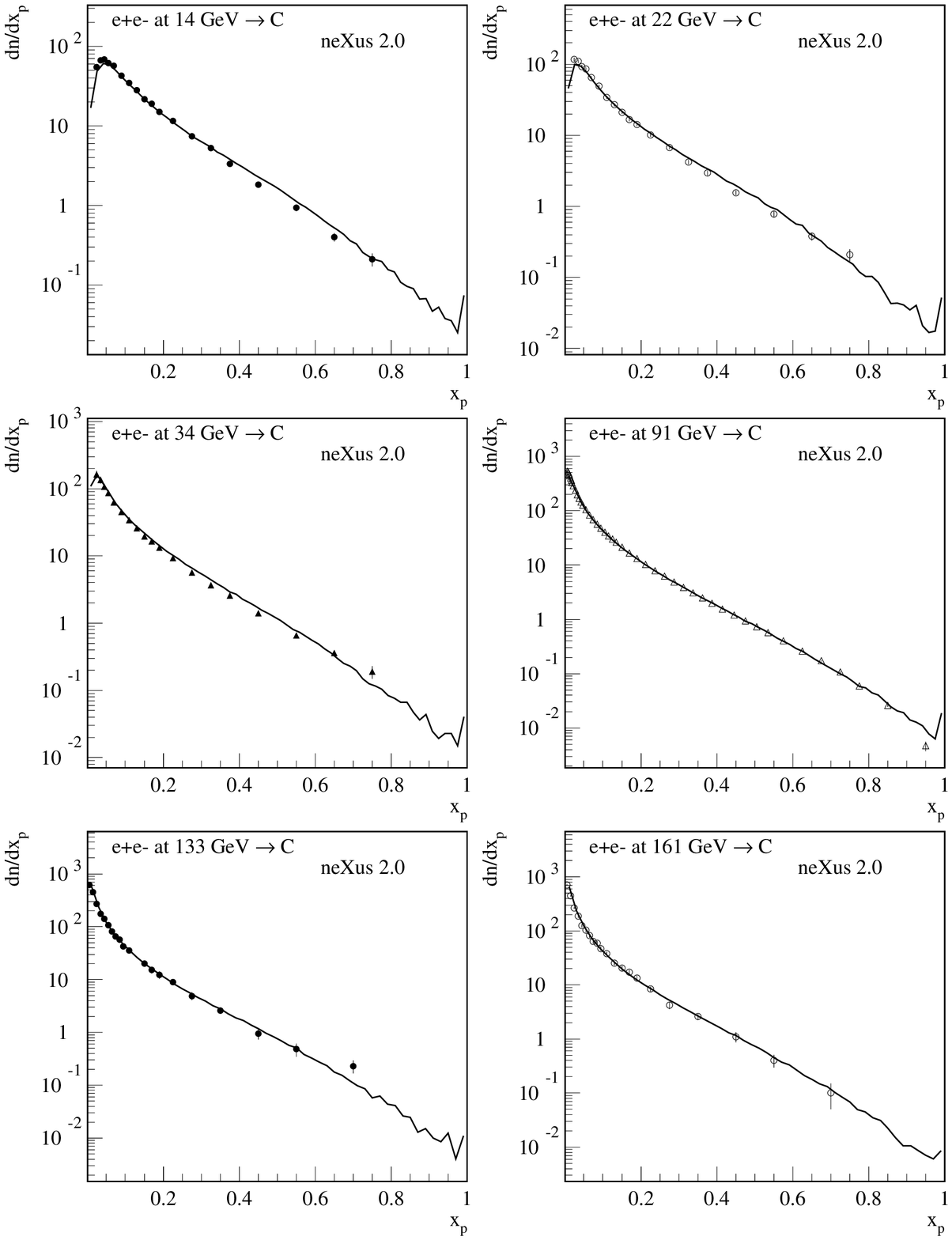}} \par}

\caption{Rapidity and longitudinal momentum distributions: data (points) and NEXUS (line).
        \label{ree2}}
\end{figure}

In real \( e^{+}e^{-} \) collisions, one has with increasing energy an increasing
probability to have kinks, which makes the plateau rising with energy, as shown
in fig. \ref{ree2}, where we show the prediction of the string model together
with experimental data from the TASSO \cite{alt84}, ALEPH \cite{bar98}, and
OPAL \cite{ale96,ack97} collaborations. We show also longitudinal momentum
fraction distributions for different energies.

\vspace{1cm}
\subsection{Hadron Flavors}

\begin{figure}[htb]
{\par\centering \resizebox*{!}{0.18\textheight}{\includegraphics{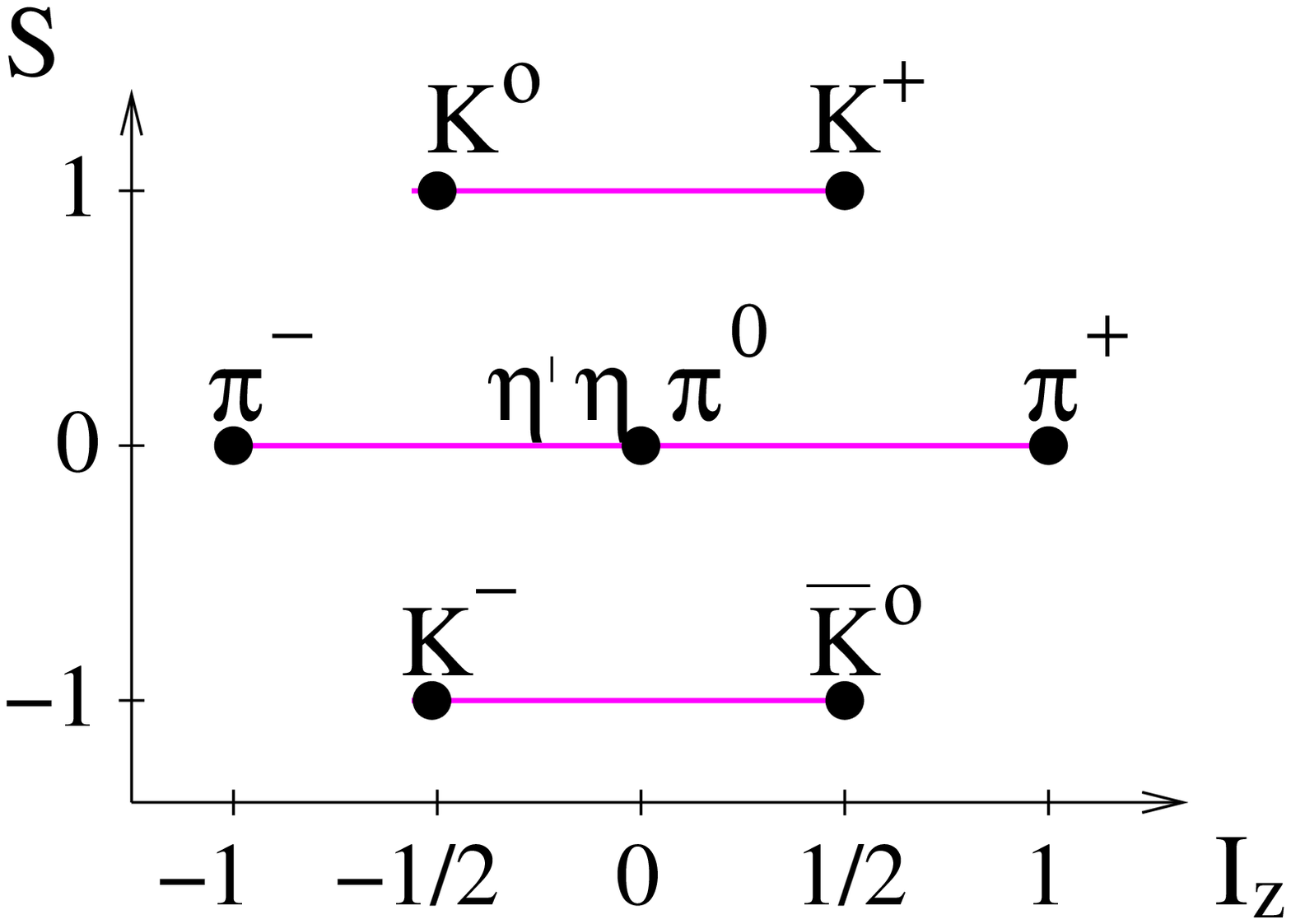}} \resizebox*{!}{0.18\textheight}{\includegraphics{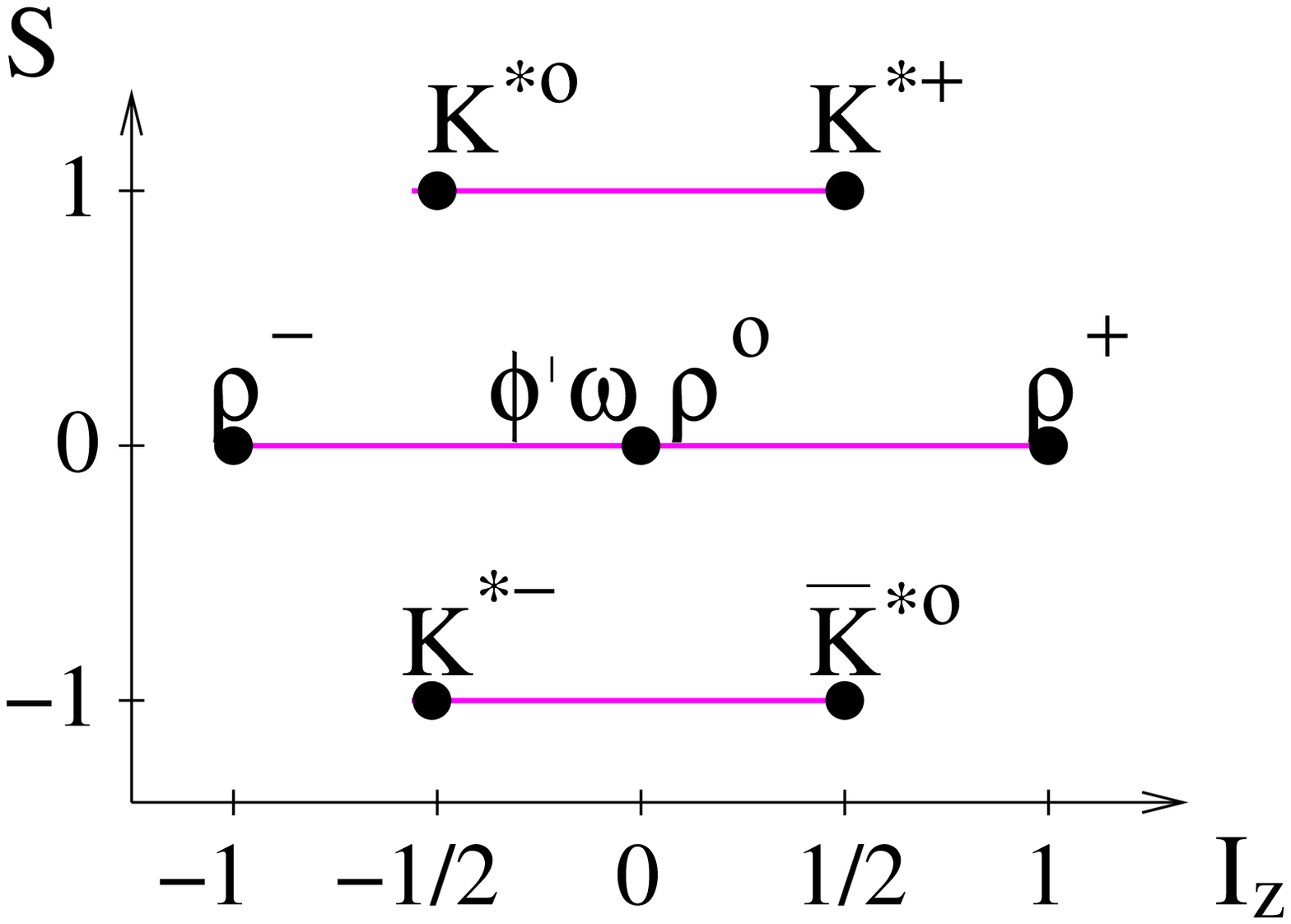}} \par}

\caption{Meson octet plus singlet.       \label{quark2b}}
\end{figure}

\begin{figure}[htb]
{\par\centering \resizebox*{!}{0.18\textheight}{\includegraphics{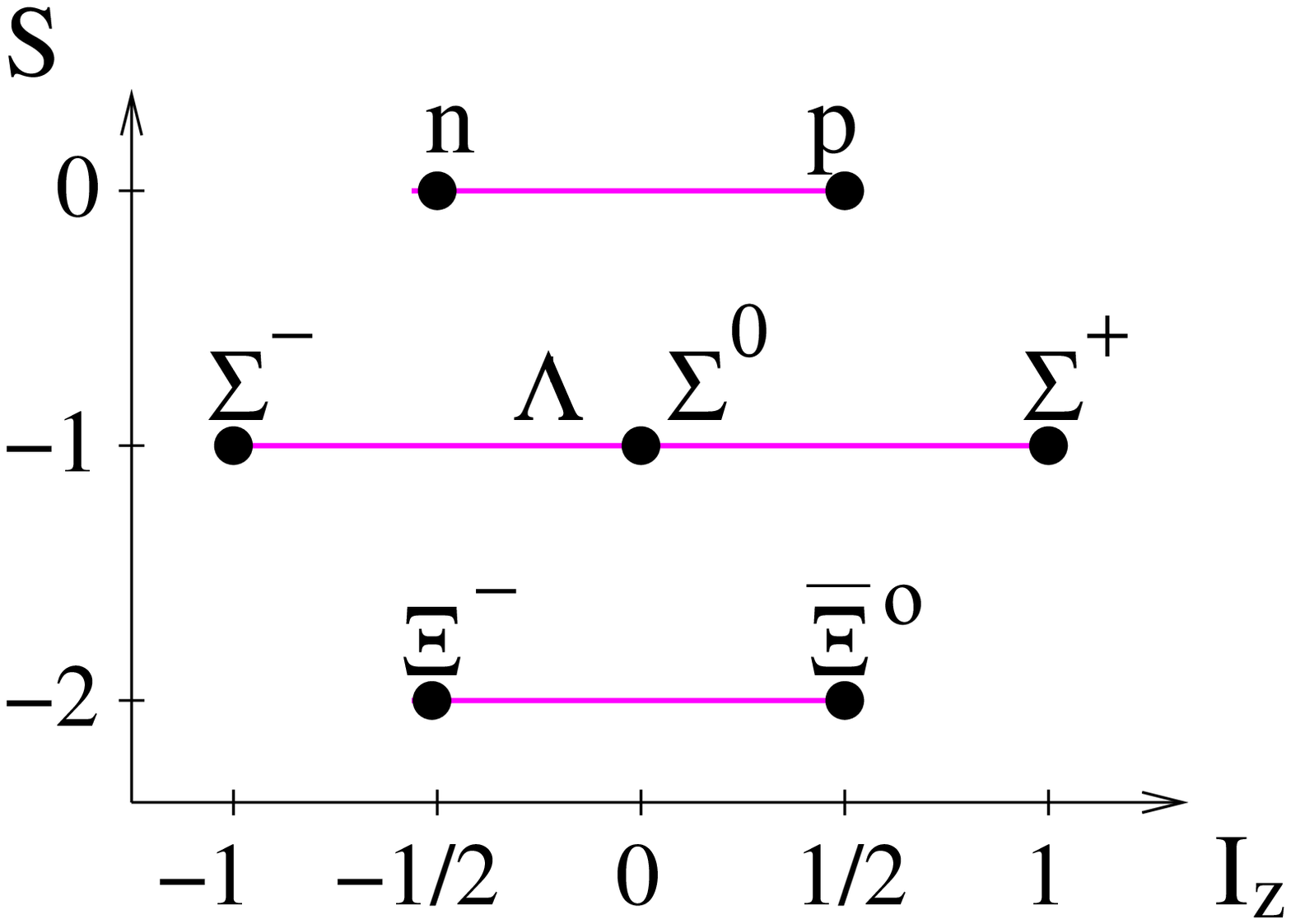}} \resizebox*{!}{0.22\textheight}{\includegraphics{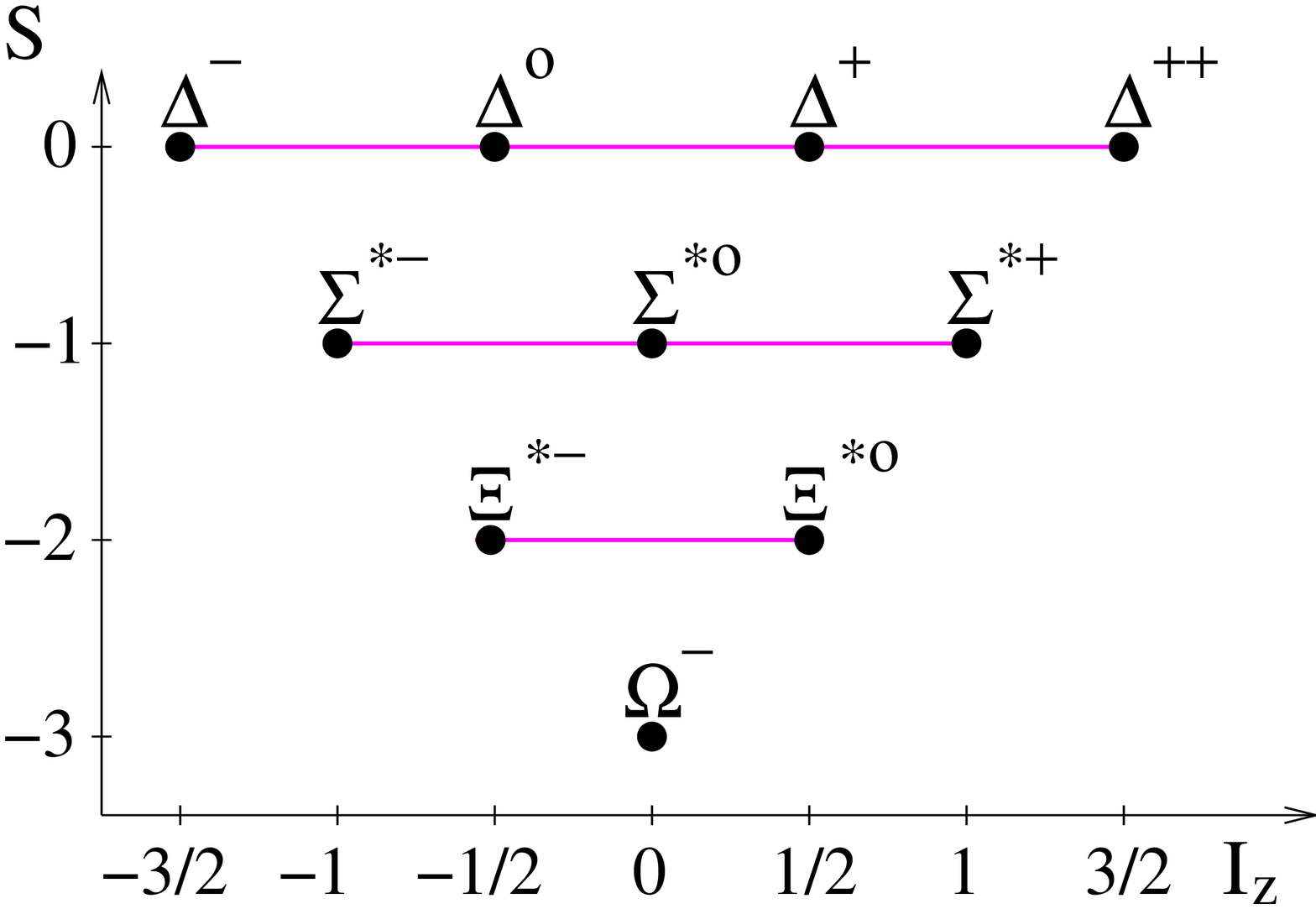}} \par}

\caption{Baryon octet and decuplet.         \label{quark1b}}
\end{figure}

There are some remarkable regularities among the hadrons, which became apparent
in the early 1960s. The first is that the baryons fall into groups of multiplicity
1, 8, 10 (singlet, octet, decuplet). The mesons come in singlets and octets.
See figs. \ref{quark2b}, \ref{quark1b}.

\begin{figure}[htb]
{\par\centering \resizebox*{!}{0.15\textheight}{\includegraphics{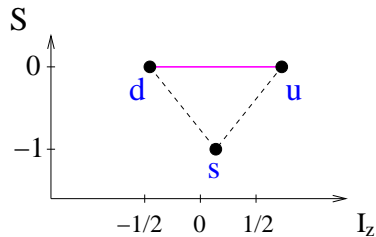}} \par}

\caption{Quark properties. \label{quark4}}
\end{figure}

\begin{figure}[htb]
{\par\centering \resizebox*{!}{0.18\textheight}{\includegraphics{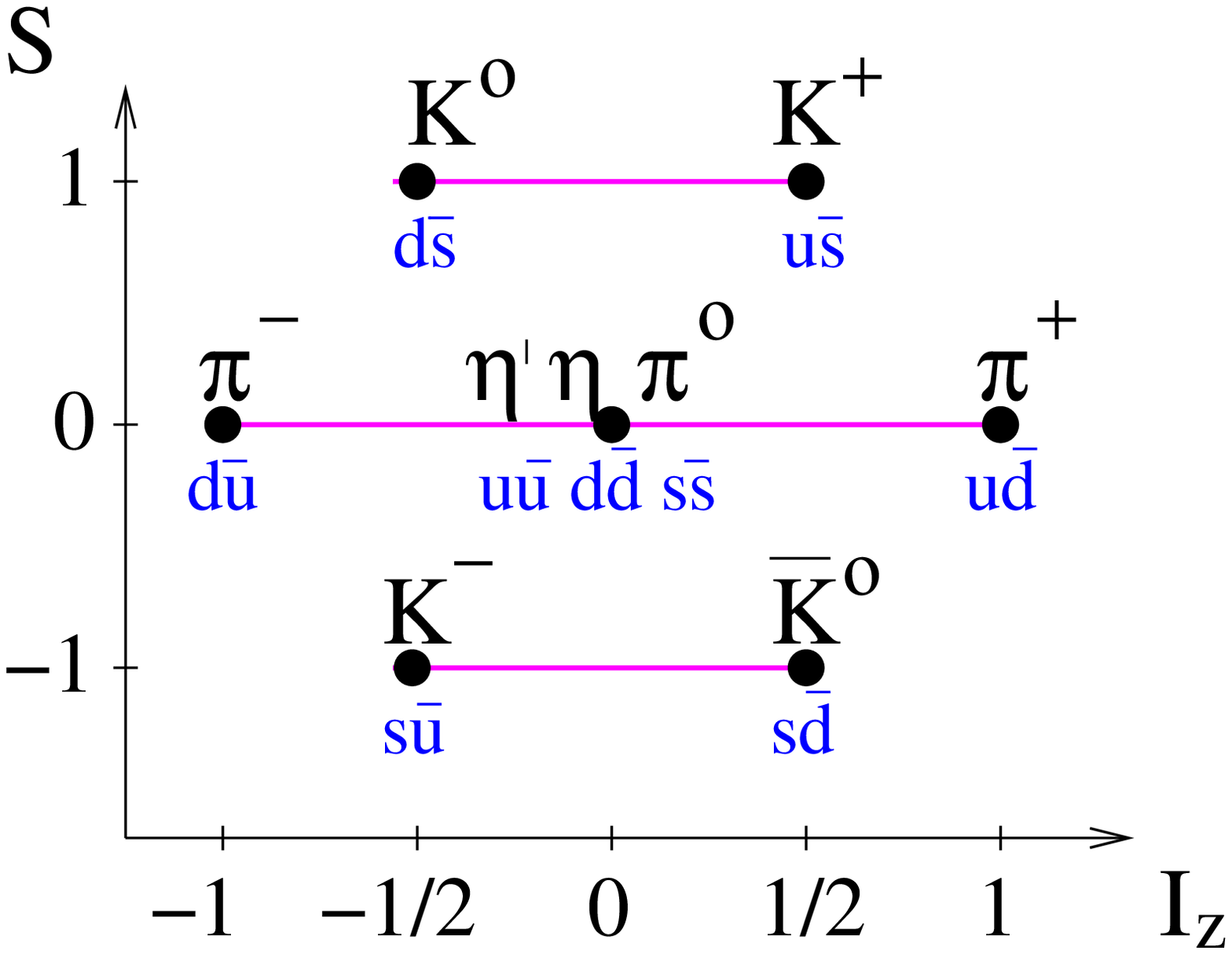}} \resizebox*{!}{0.18\textheight}{\includegraphics{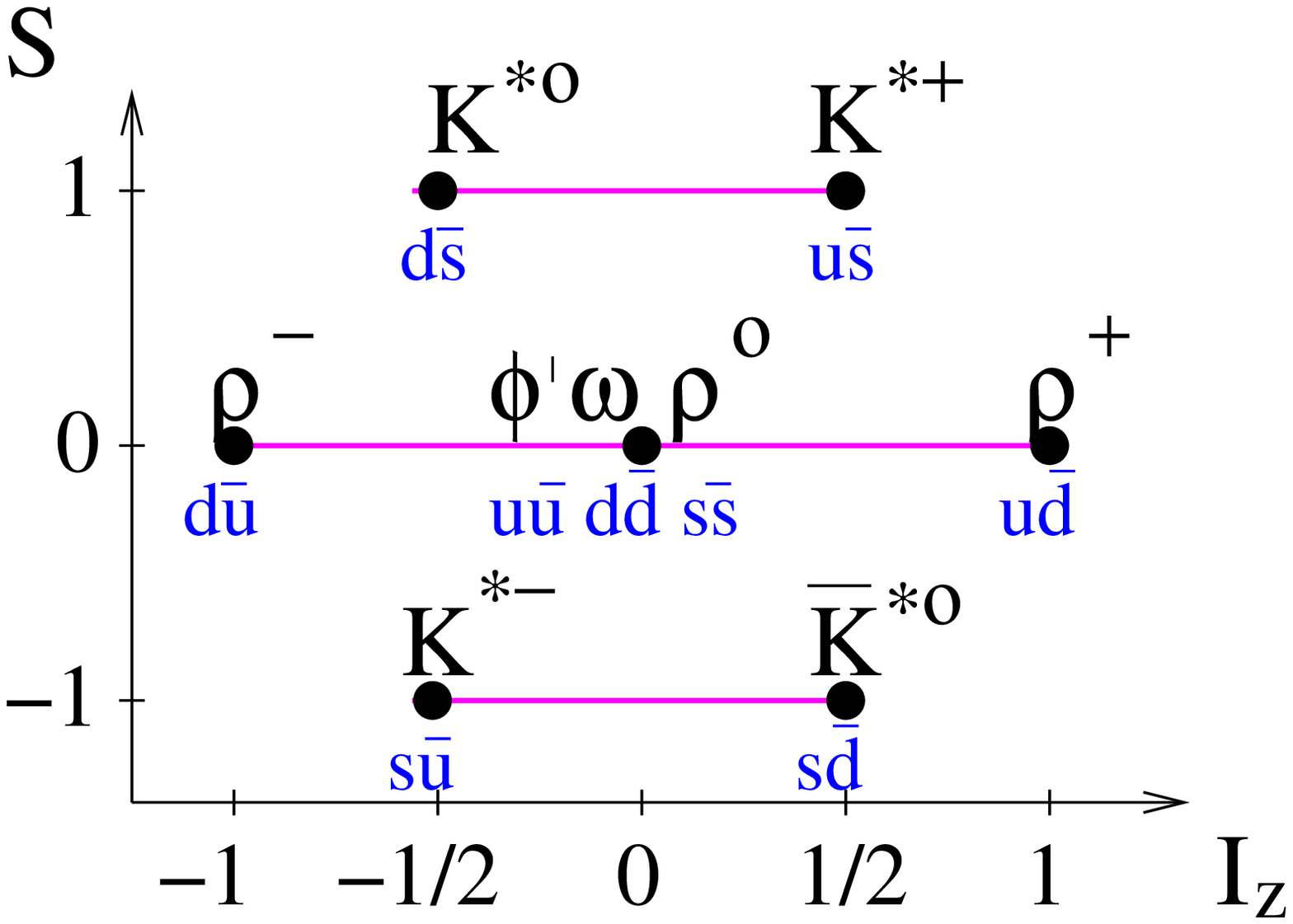}} \par}

\caption{The quark content of mesons.\label{quark6b}}
\end{figure}

\begin{figure}[htb]
{\par\centering \resizebox*{!}{0.18\textheight}{\includegraphics{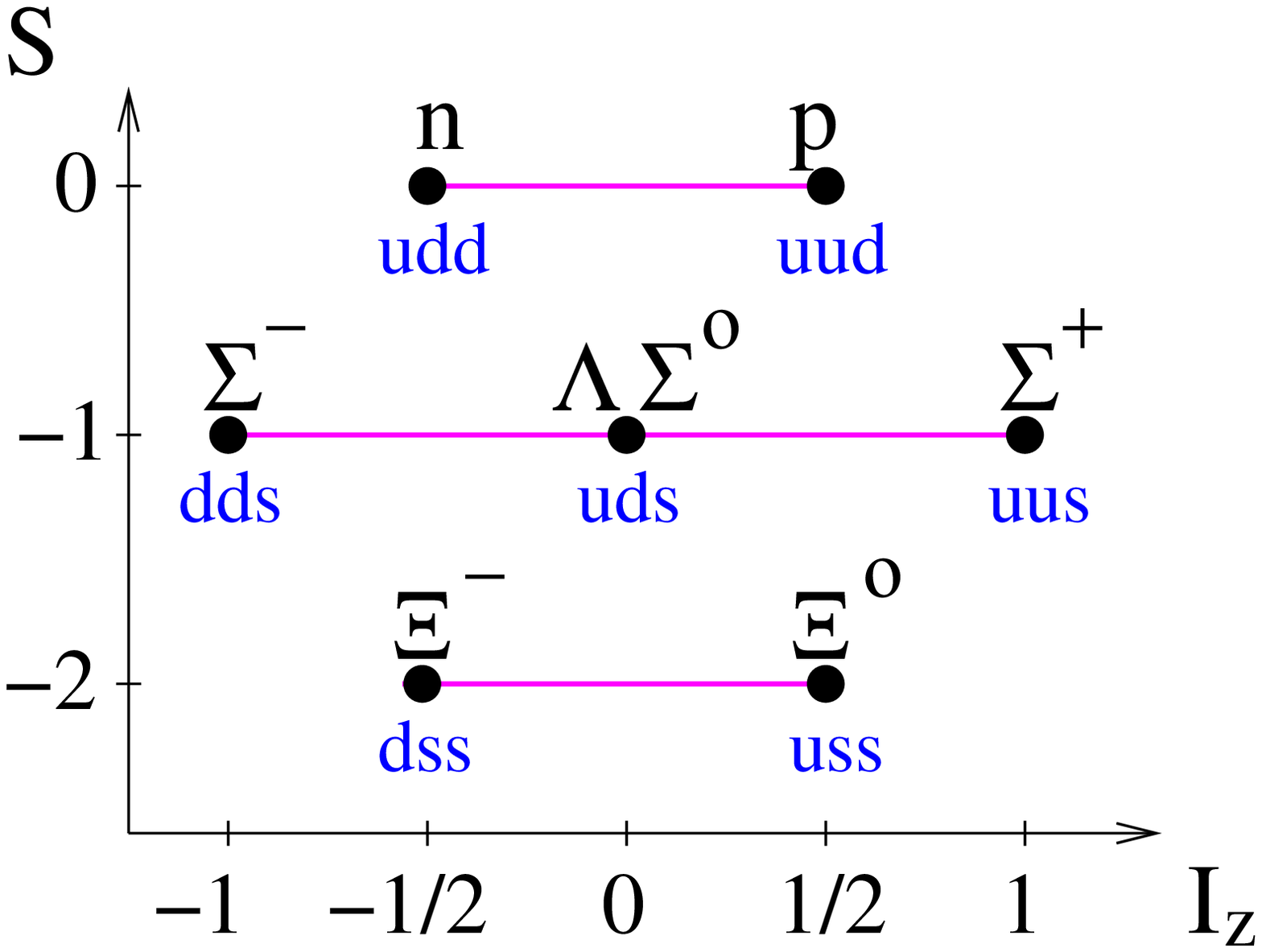}} \resizebox*{!}{0.23\textheight}{\includegraphics{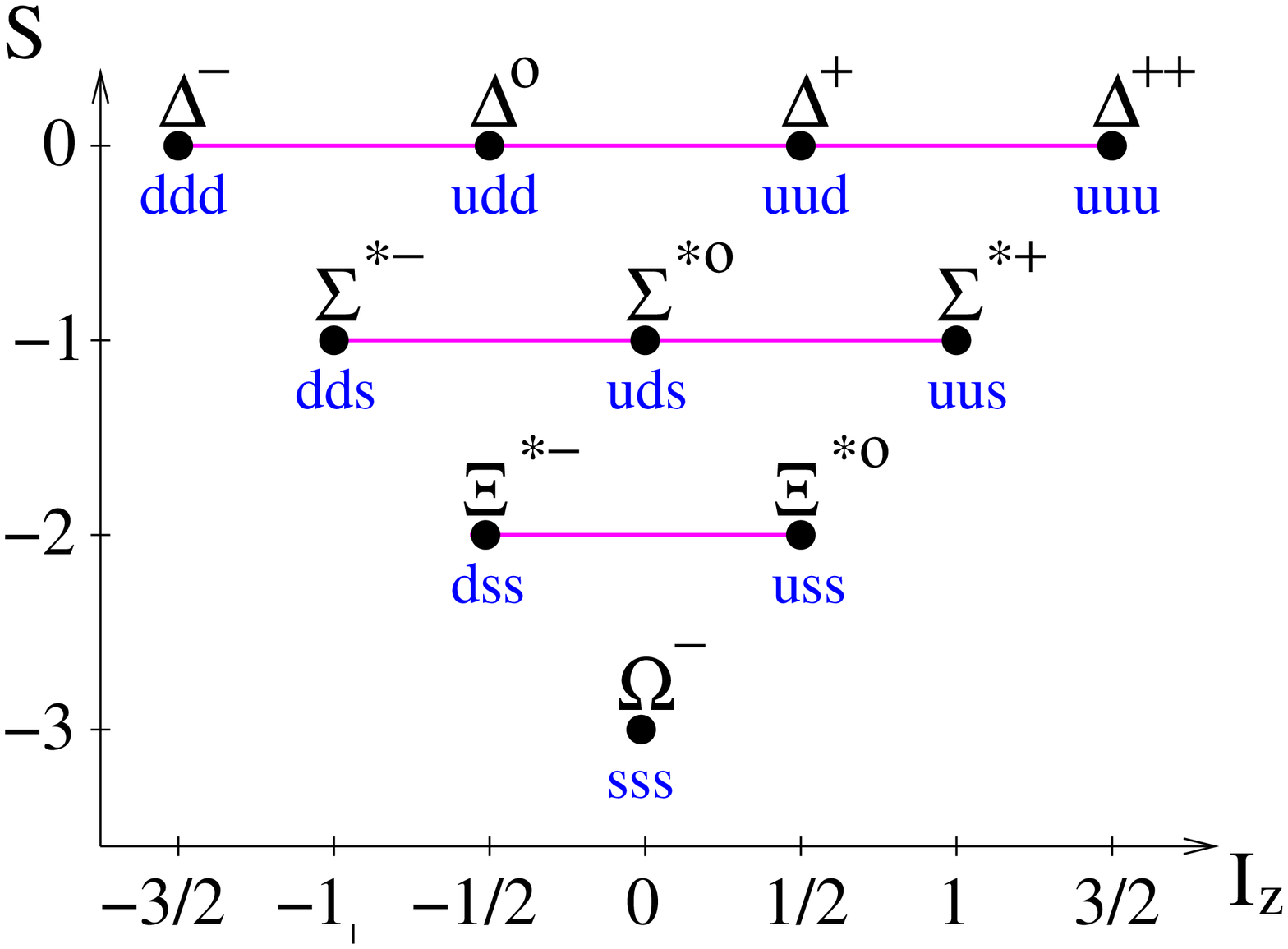}} \par}

\caption{The quark content of baryons.\label{quark5b}}
\end{figure}

These structures can be understood in the quark model for hadrons: the baryons
are composed of three quarks, the antibaryons of three antiquarks and the mesons
of a quark plus an antiquark. There are six flavors of quarks, the three most
abundant ones being the \( u \), \( d \), and \( s \) flavor, the properties
being given in table \ref{prop}, shown in fig. \ref{quark4}.
\begin{table}[htb]
{\centering \begin{tabular}{|c|c|c|c|}
\hline 
name&
u&
d&
s\\
\hline 
\hline 
charge&
2/3&
-1/3&
-1/3\\
\hline 
strangeness&
0&
0&
-1\\
\hline 
isospin&
1/2&
-1/2&
0\\
\hline 
\end{tabular}\par}

\caption{Quark properties. \label{prop}}
\end{table}
 In figs. \ref{quark6b} and \ref{quark5b}, we indicate the quark content of
the most frequent mesons and baryons. So the quark model can easily explain
the striking regularities of the hadrons. 
 
\begin{figure}[htb]
{\par\centering \resizebox*{!}{0.06\textheight}{\includegraphics{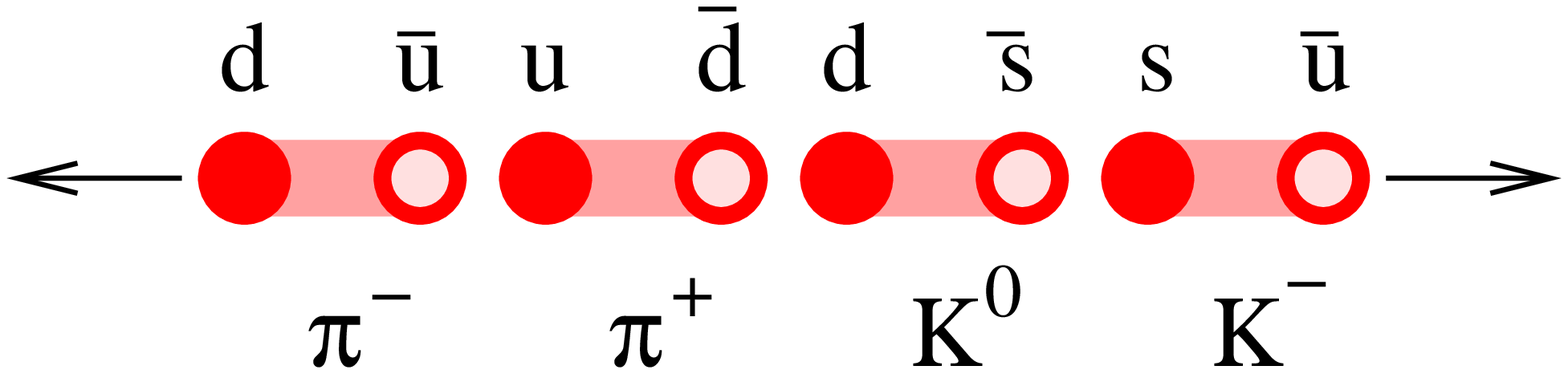}} 
\( \qquad  \)\resizebox*{!}{0.055\textheight}{\includegraphics{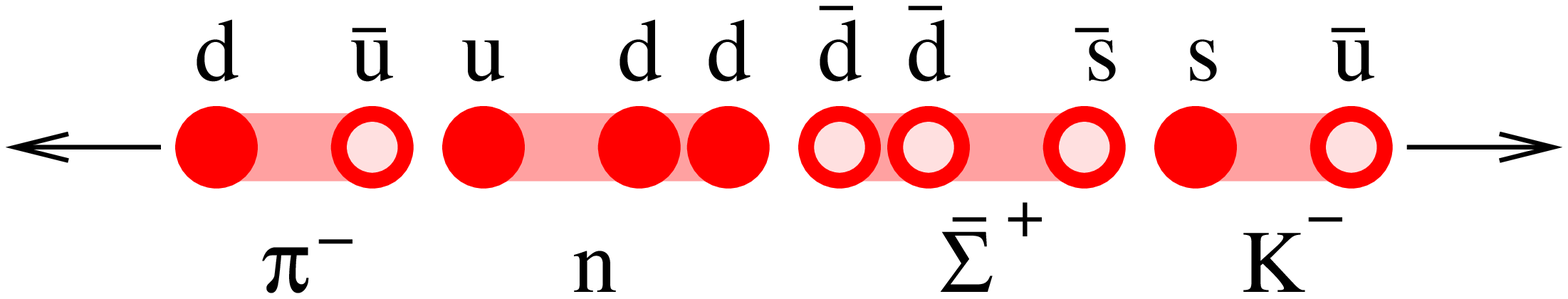}} \par}

\caption{Hadron production from a quark-antiquark string.         \label{ee15}}
\end{figure}

\begin{figure}[htb]
{\par\centering \resizebox*{!}{0.42\textheight}{\includegraphics{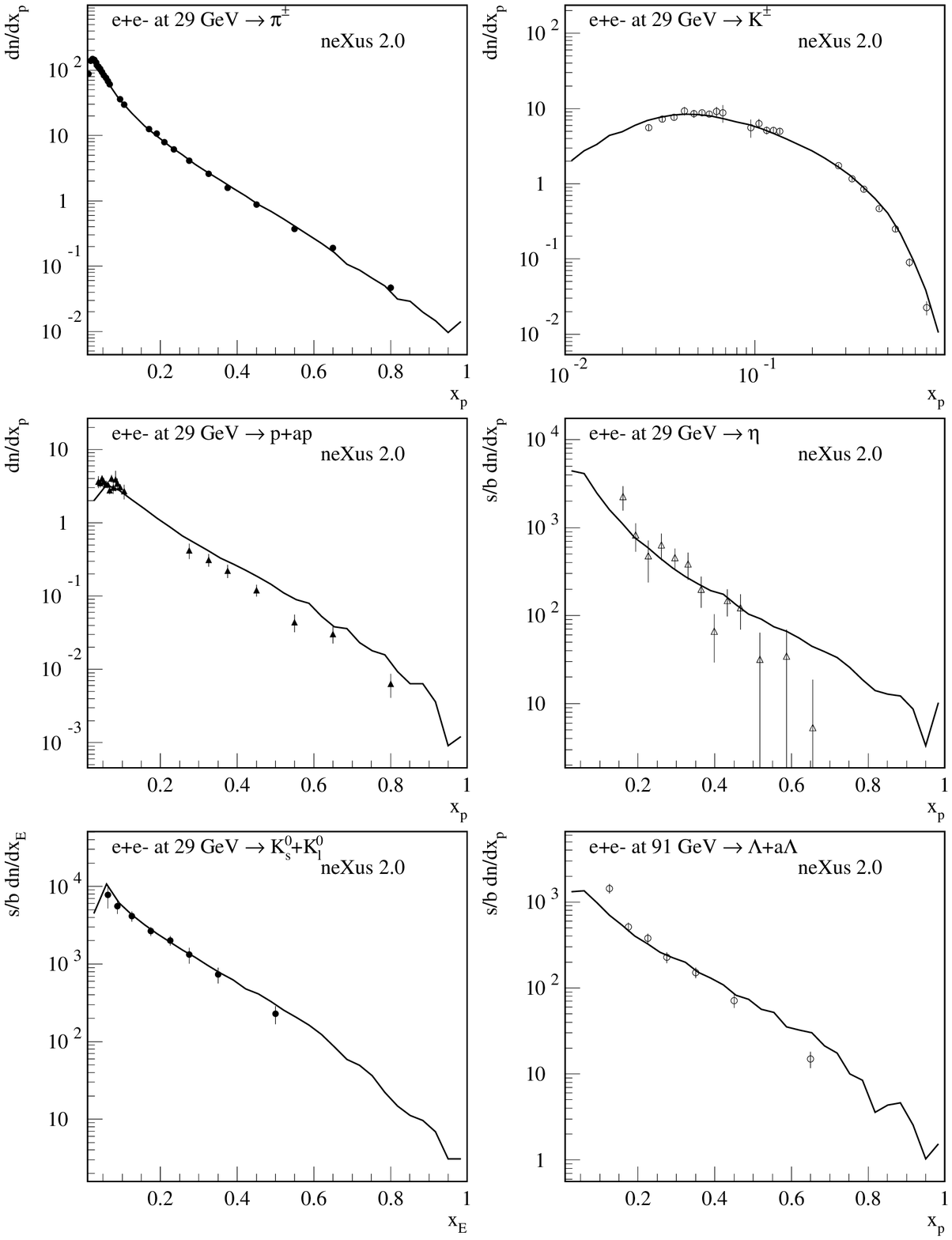}} \resizebox*{!}{0.42\textheight}{\includegraphics{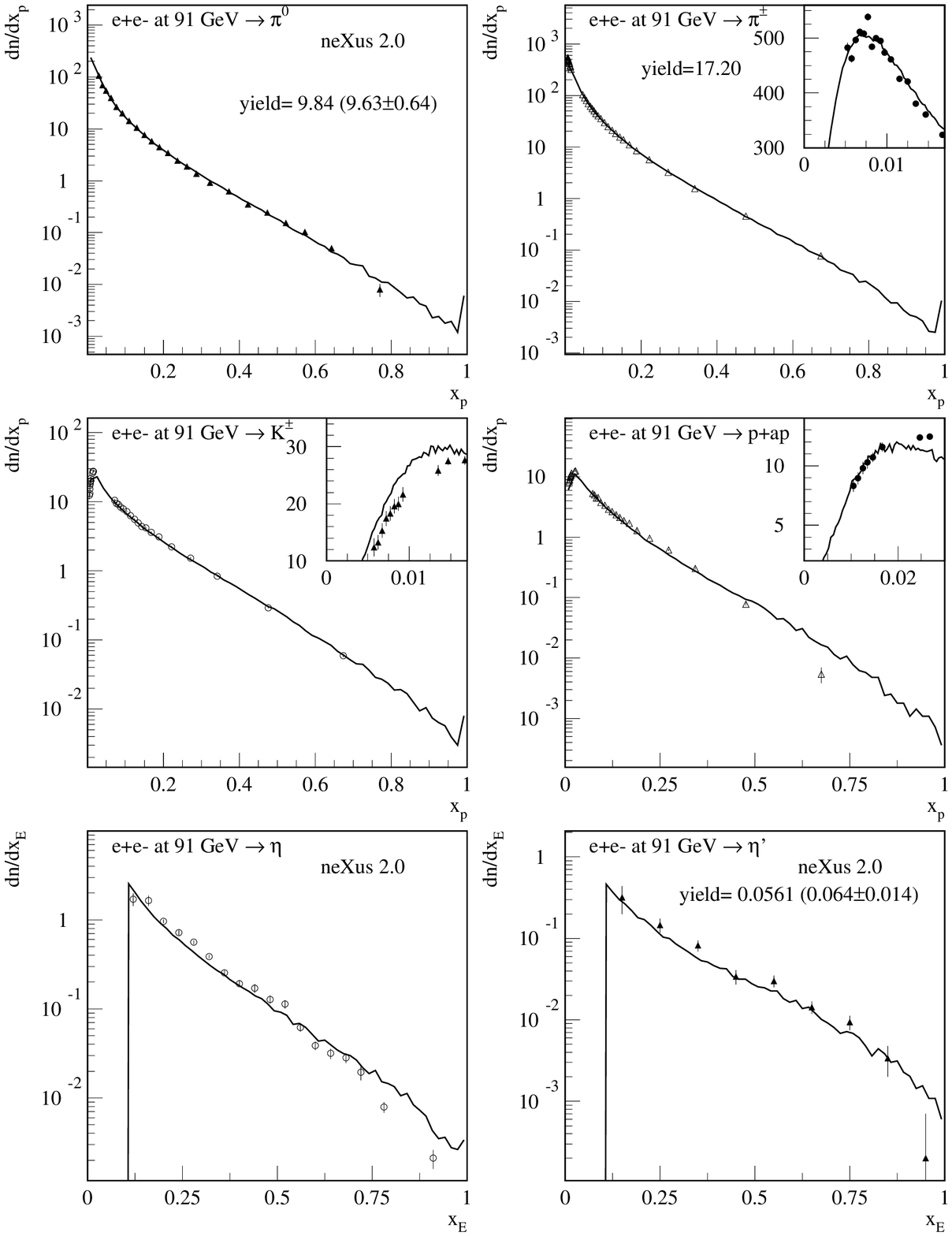}} \par}

\caption{Hadron production in \protect\( e^{+}e^{-}\protect \) annihilation at 29 and
91 GeV.         \label{ee29}}
\end{figure}

 It will be the basis of all models of string fragmentation, as shown in fig.
\ref{ee15}. A string break is realized via quark-antiquark production, such
that the quark-antiquark pair screens the color field. The string fragments
consist then of quark-antiquark pairs, they are therefore mesons. It is also
possible that the string breaks via diquark-antidiquark production, which amounts
to baryon and antibaryon creation.In fig. \ref{ee29}, we show some results
concerning the production of identified hadrons in electron-positron annihilation.
One observes that the string model works to a high precision.
 
\vspace{1cm}
\section{Parton Model}

Whereas leptons are point-like in their behavior, it is not inconceivable that
the quarks too enjoy this property. If we think of the hadrons as complicated
``atoms'' or ``molecules'' of quarks, then at high energies and momentum
transfers, where we are probing the inner structure, we may discover a simple
situation, with the behavior controlled by almost free, point-like constituents.
The idea that hadrons possess a ``granular'' structure and that the ``granules''
behave as hard point-like, almost free (but nevertheless confined) objects,
is the basis of Feynmans (1969) parton model.

The essence of the parton model is the assumption that, when a sufficiently
high momentum transfer reaction takes place, the projectile, be it a lepton
or a parton inside a hadron, sees the target as made up of almost free constituents,
and is scattered by a single, free, effectively massless constituent.

\subsection{Deep Inelastic Scattering }

\begin{figure}[htb]
{\par\centering \resizebox*{!}{0.08\textheight}{\includegraphics{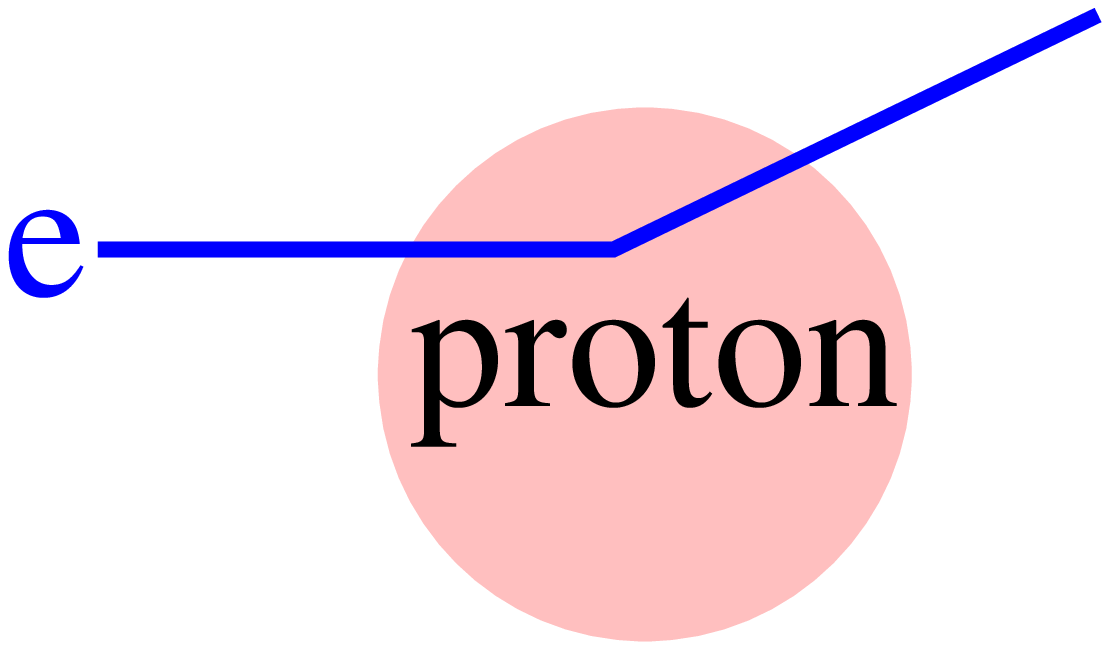}} \( \qquad  \)\resizebox*{!}{0.08\textheight}{\includegraphics{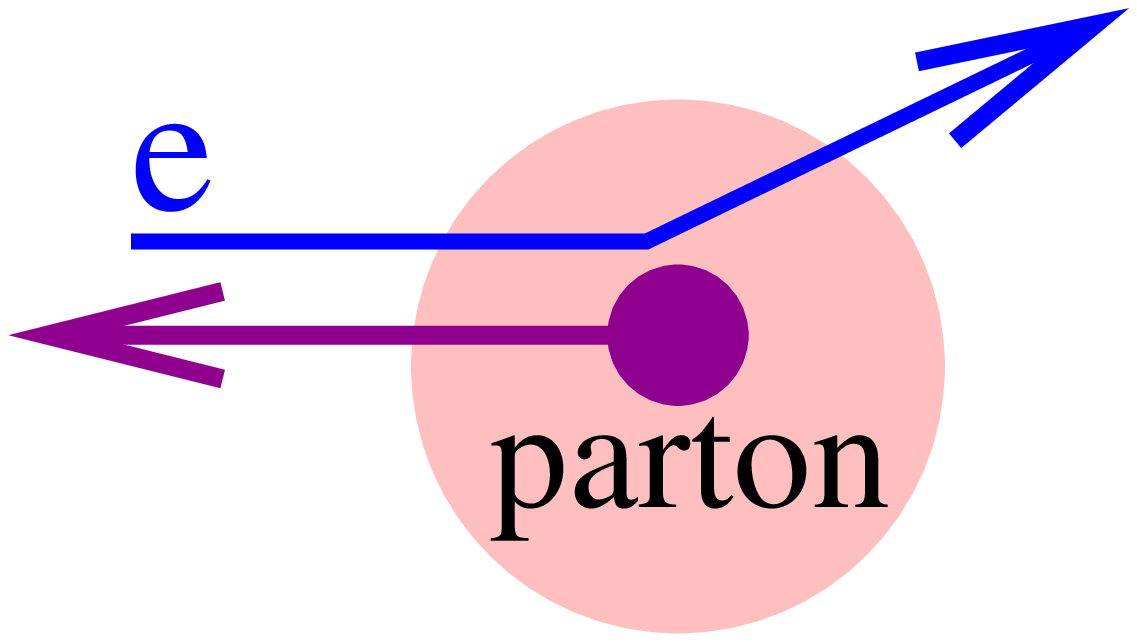}} \par}

\caption{lepton-proton scattering.\label{dis1}}
\end{figure}

\begin{figure}[htb]
{\par\centering \resizebox*{!}{0.2\textheight}{\includegraphics{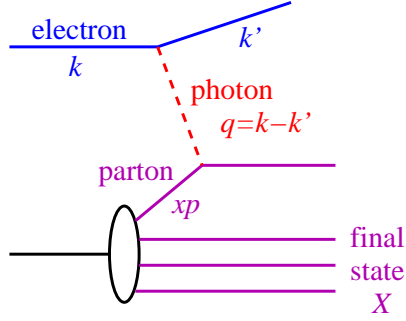}} \par}

\caption{One photon exchange in a \protect\( ep\protect \) reaction.         \label{dis3}}
\end{figure}

\begin{figure}[htb]
{\par\centering \resizebox*{!}{0.08\textheight}{\includegraphics{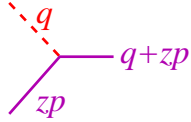}} \par}

\caption{The photon-parton vertex.         \label{dis4}}
\end{figure}

The historical way to study the hadronic structure was using point-like leptons
as projectiles hitting a proton target. There is a basic difference compared
to \( e^{+}e^{-} \) scattering: the proton is a composite particle, \( e^{+} \),
\( e^{-} \) are elementary particles. so one can probe the internal structure
of the proton, see fig. \ref{dis1}(left). 
In a lepton-proton scattering, one can measure the momentum distributions of
constituents (partons) see fig. \ref{dis1}(right). In lowest order of perturbation
theory, the reaction is described by one photon exchange diagram, see fig. \ref{dis3}.
Here \( k \) is known and \( k' \) is measured, so the momentum transfer \( q \)
is known.
 One studies the cross section as function of two variables: the photon virtuality
\( Q^{2}=-q^{2} \) and the Bjorken variable \( x=Q^{2}/2pq. \) Why \( Q^{2} \)?
Because \( Q^{2} \) sets the resolution scale: \( \Delta x=1/Q^{2} \). The
bigger \( Q^{2} \) the deeper one looks into the proton. Why x? Consider a
parton with a momentum fraction \( z \) (momentum \( zp \)), see fig. \ref{dis4}. 
In the reaction it received the transferred momentum \( q \), so its new momentum
is \( q+zp \). But the parton is massless, so \( q^{2}+2qzp+z^{2}p^{2}=0 \),
and therefore \( z=x \). A reaction with a certain value of \( x \) probes
a parton with momentum fraction \( z=x \), which means that parton momentum
distribution are measurable.

\begin{figure}[htb]
{\par\centering \resizebox*{!}{0.15\textheight}{\includegraphics{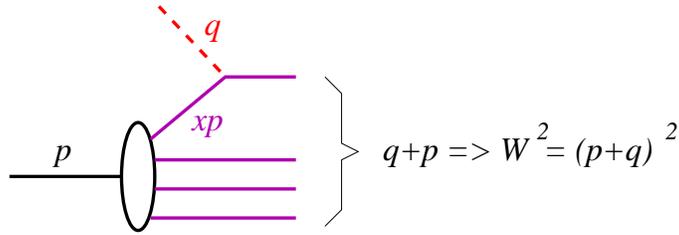}} \par}

\caption{The total mass \protect\( W\protect \) of the hadronic final state.         \label{dis5}}
\end{figure}

Let us do some kinematics: the virtual photon transfer being \( q \) and the
initial proton momentum being \( p \), the final state mass \( W \) is given
as \( W^{2}=(p+q)^{2} \), see fig. \ref{dis5}. We have
 ,

\[
W^{2}=(q+p)^{2}=q^{2}+2pq+p^{2}\approx -Q^{2}+Q^{2}/x,\]
which gives

\[
W^{2}=Q^{2}(\frac{1}{x}-1)\]
So we arrive at an interesting result: small \( x \) corresponds to a big final
state mass \( W \).

\begin{figure}[htb]
{\par\centering \resizebox*{!}{0.12\textheight}{\includegraphics{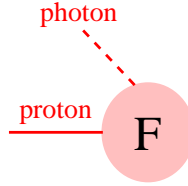}} \par}

\caption{Proton structure functions.         \label{dis6}}
\end{figure}

One can write the \( ep \) cross section as 

\[
\frac{d\sigma ^{ep}}{dQ^{2}dx}=\frac{\alpha }{\pi Q^{2}x}\left[ \frac{1+(1-y)^{2}}{2}\sigma _{T}^{\gamma p}+(1-y)\sigma _{L}^{\gamma p}\right] ,\]
 with \( y=pq/pk \), and

\[
\sigma _{T}^{\gamma p}=\frac{4\pi ^{2}\alpha }{Q^{2}}(F_{2}-F_{L}),\sigma _{L}^{\gamma p}=\frac{4\pi ^{2}\alpha }{Q^{2}}F_{L},F_{L}\ll F_{2}.\]
 \( F_{2} \) and \( F_{L} \) describe the proton structure as seen by the
virtual photon, see fig. \ref{dis6}.

A first look reveals \( F_{2}(x,Q^{2}) \)to be only a function of x (scaling).
This can be explained within the naive Parton Model, where the proton is considered
to be a incoherent sum of partons (quarks) of flavor \( i \), which are distributed
according to parton distribution functions \( f_{i} \) , so
\[
F_{2}(x,Q^{2})=\sum _{i}e_{i}\, x\, f_{i}(x).\]
\begin{figure}[htb]
{\par\centering \resizebox*{!}{0.12\textheight}{\includegraphics{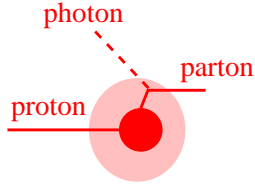}} \par}

\caption{Parton model.         \label{dis7}}
\end{figure}

\begin{figure}[htb]
{\par\centering \resizebox*{!}{0.24\textheight}{\includegraphics{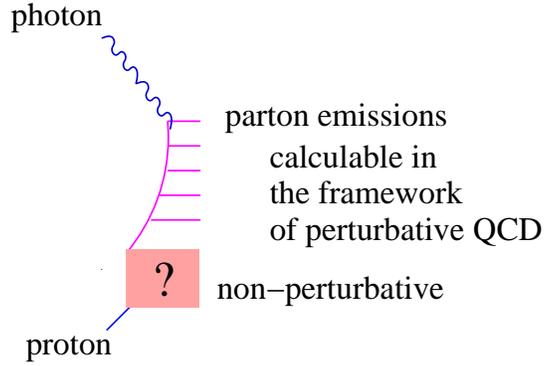}} \par}

\caption{Successive parton emissions.         \label{dis9}}
\end{figure}

Looking more carefully, on finds that \( F_{2}(x,Q^{2}) \) depends slightly
on \( Q^{2} \) (scale breaking). The partons are still distributed according
to parton distribution functions \( f_{i}(x,Q^{2}) \), which are now depending
on a scale (defined by the probe). And we still have
\[
F_{2}(x,Q^{2})=\Sigma _{i}e_{i}xf_{i}(x,Q^{2}).\]
 This formula takes in account the successive emissions of virtual partons,
carrying less and less momentum. The photon virtualities, \( Q^{2} \), are
ordered down to some minimum value (this part is calculable in the framework
of perturbative QCD). Below this minimum value, we have soft physics (non-perturbative
regime), see fig. \ref{dis9}.

\begin{figure}[htb]
{\par\centering \resizebox*{!}{0.2\textheight}{\includegraphics{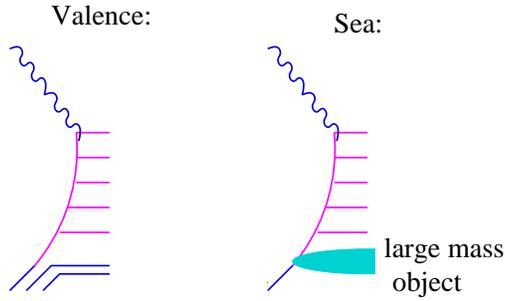}} \par}

\caption{Two contributions for the proton structure function.         \label{dis8}}
\end{figure}

\begin{figure}[htb]
{\par\centering \resizebox*{!}{0.28\textheight}{\includegraphics{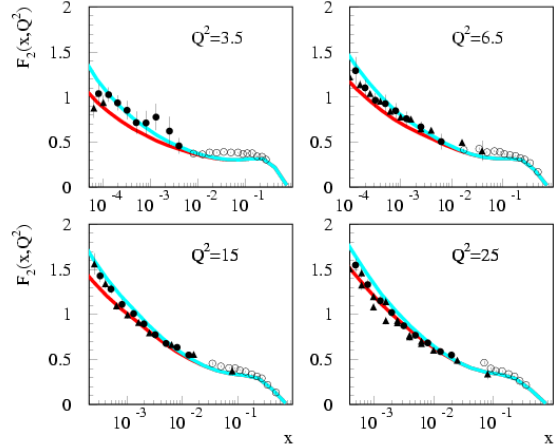}} \par}

\caption{The structure function \protect\( F_{2}\protect \). \label{f2s}}
\end{figure}

Let us consider the emission of the first (softest) parton being emitted from
the non-perturbative area. If the parton carries a fraction \( x \) of the
momentum of the proton, the mass of the soft object ''proton minus parton''
has a mass given by \( m^{2}=Q_{0}^{2}/x \), where \( Q_{0} \) is a typical
soft virtuality (of the order 1 GeV). This has interesting consequences: in
case of sea quarks with distributions of the form \( 1/x \), one has typically
small \( x \) and therefore a large mass \( m \). For valence quarks, on the
other hand, a \( 1/\sqrt{x} \) distribution provides in general large \( x \)
values and therefore a small mass \( m \). This means that in case of sea quarks,
there is a large mass and small virtuality object between the proton and the
parton, and we therefore have to consider two contributions to the structure
functions, as indicated in fig. \ref{dis8}. The large mass object is considered
to be a Pomeron, to be discussed in detail later. 
The valence contribution provides a peak at large \( x \) and drops fast for
small values of \( x \), whereas the sea contribution is important at small
\( x \) and drops very fast towards large \( x \). The sum of the two is shown
in fig. \ref{f2s}.

\subsection{The Parton Model for pp}

\begin{figure}[htb]
{\par\centering \resizebox*{!}{0.2\textheight}{\includegraphics{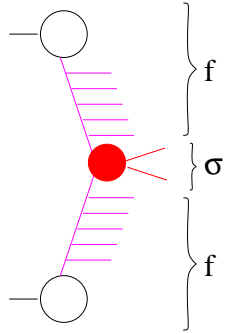}} \par}

\caption{The inclusive cross section for producing partons. \label{par3}}
\end{figure}

For \( pp \) interactions one uses the same concept as for lepton-proton scattering.
Each proton contains partons distributed as \( f_{i}(x,Q^{2}) \). Two of the
partons (one from the first proton and the other from the second proton) interact
via elementary interactions. The inclusive cross section for producing a parton
k is

\[
\frac{d\sigma ^{pp\rightarrow k}}{dyd^{2}p_{t}}=\sum _{i,j}\int dx_{1}dx_{2}f_{i}(x_{1},Q^{2})f_{j}(x_{2},Q^{2})\frac{d\hat{\sigma }^{ij\rightarrow k}}{dyd^{2}p_{t}},\]
 (see fig. \ref{par3}), where \( f_{i}(x,Q^{2}) \) are the perturbative parton
densities, measured by performing global fits of data taken from large sets
of experiments as lepton-nucleon deep inelastic scattering and others. \( d\hat{\sigma }/dyd^{2}p_{t} \)
are partonic cross-section for the hard processes, calculable in perturbation
theory. 
Here, one assumes universality of parton densities (independent of the hard
processes, for large \( Q^{2} \)) and factorization (the possibility to separate
the parton density functions from the partonic cross section). The assumption
is based in the fact that hard processes (\( Q\gg m_{p} \)) occur in very short
time (\( \tau \sim 1/Q \)), much lesser than the typical interaction times
for the binding of the proton (hadron) (\( \tau \sim 1/m_{p} \)). As a result,
to study inclusive processes at large \( Q^{2} \) it is sufficient to consider
the interactions between the external probe and a single parton.

\begin{figure}[htb]
{\par\centering \resizebox*{!}{0.15\textheight}{\includegraphics{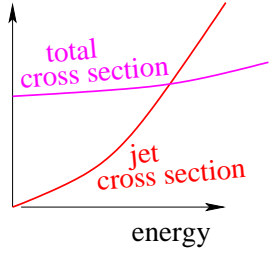}} \par}

\caption{The jet cross section grows very fast with energy .        \label{par4}}
\end{figure}

The model works well for most of the high energy physics experiments, but diverges
for small transverse momentum. Why? Because soft (non-perturbative) physics
enters. A solution is to introduce some lower limit (cutoff \( p_{0} \)) for
the transverse momentum. Integrating over rapidity and the transverse momentum
above the cutoff, we get the jet cross section \( \sigma (p_{0}) \). Now another
problem appears: the jet cross section grows very fast with energy, becoming
finally bigger than the total one, see fig. \ref{par4}.
The real solution amounts to studying multiple scattering. The jet cross section
is an inclusive one: several jets may contribute. So the parton model is very
useful but is not the whole history. One needs a multiple scattering theory.

There is currently much discussion about saturation. What does it mean? Consider
partons with transverse momentum \( p_{0} \). Each one occupies a transverse
area \( \pi /p_{0}^{2} \), whereas the transverse area of the nucleon is \( \pi R_{A}^{2} \).
If the number \( N_{A}(s,p_{0}^{2}) \) of partons is sufficiently high, they
fill completely the transverse area of the nucleus. The relation
\[
N_{A}(s,p_{0}^{2})\pi /p_{0}^{2}=\pi R_{A}^{2}\: \mathrm{or}\: p_{0}^{2}=N_{A}(s,p_{0}^{2})/R_{A}^{2}\]
 defines therefore the so-called saturation scale \( p_{0}^{2} \). Since \( N_{A}(s,p_{0}^{2}) \)
increases with \( s \) and decreases with \( p_{0}^{2} \), then the condition
\( p_{0}^{2}=N_{A}(s,p_{0}^{2})/R_{A}^{2} \) is solved by a function \( p_{0}^{2}(s) \)
which increases with s, such that at sufficiently high energy the scale \( p_{0}^{2}(s) \)
in in the perturbative domain.

\section{Multiple Scattering Theory}

\subsection{Reminder: some Elementary Quantum Mechanics}

\begin{figure}[htb]
{\par\centering \resizebox*{!}{0.12\textheight}{\includegraphics{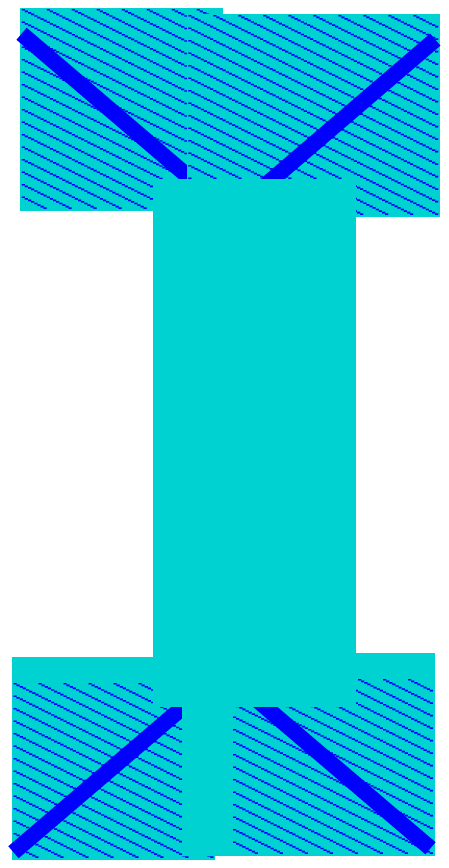}} \( \qquad  \)\resizebox*{!}{0.12\textheight}{\includegraphics{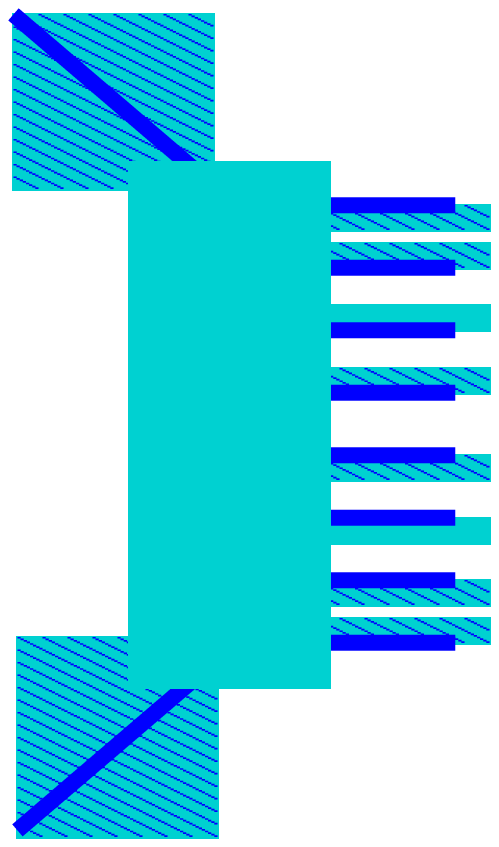}} \par}

\caption{An elastic scattering amplitude \protect\( T_{2\rightarrow 2}\protect \) (left)
and an inelastic amplitude \protect\( T_{2\rightarrow X}\protect \) (right).\label{t0}}
\end{figure}

\begin{figure}[htb]
{\par\centering \resizebox*{!}{0.12\textheight}{\includegraphics{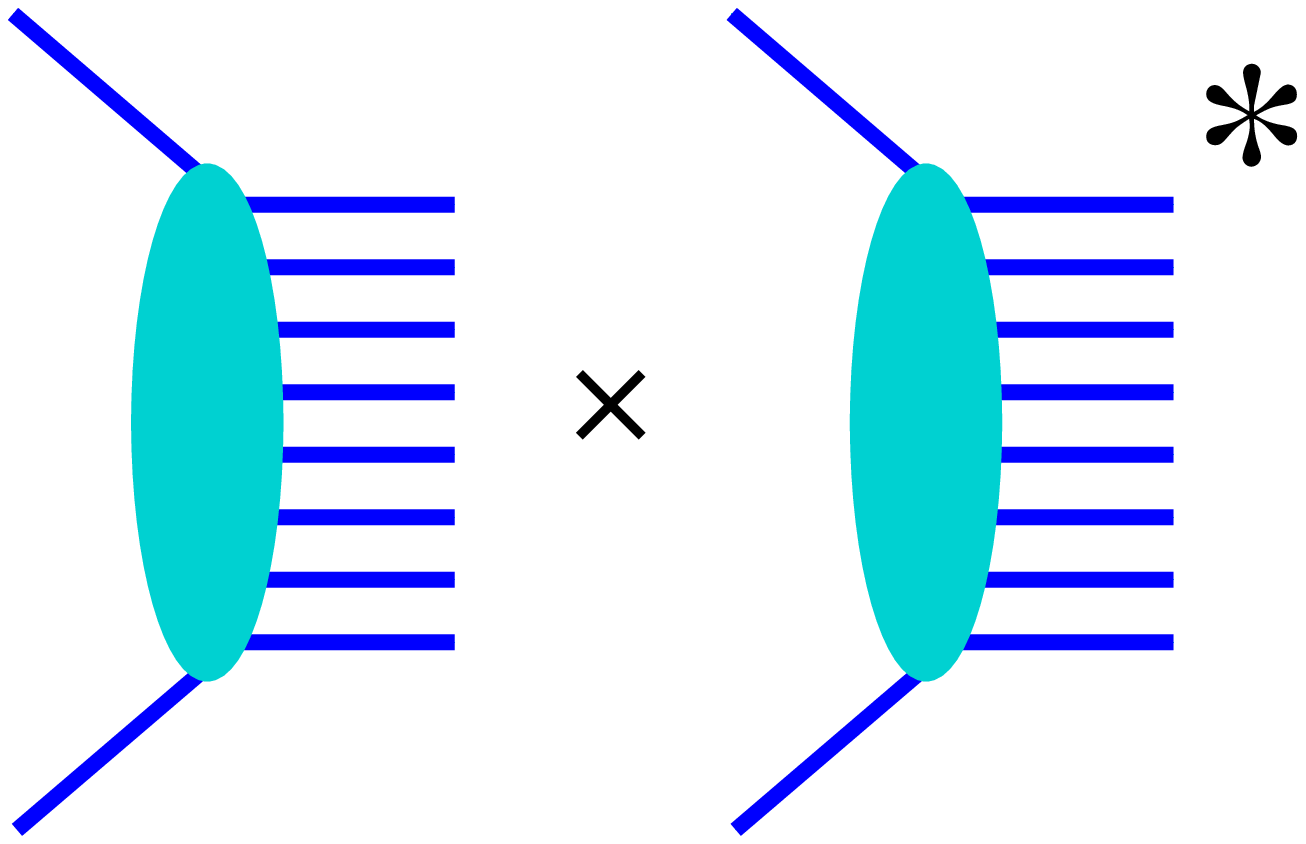}} \resizebox*{!}{0.12\textheight}{\includegraphics{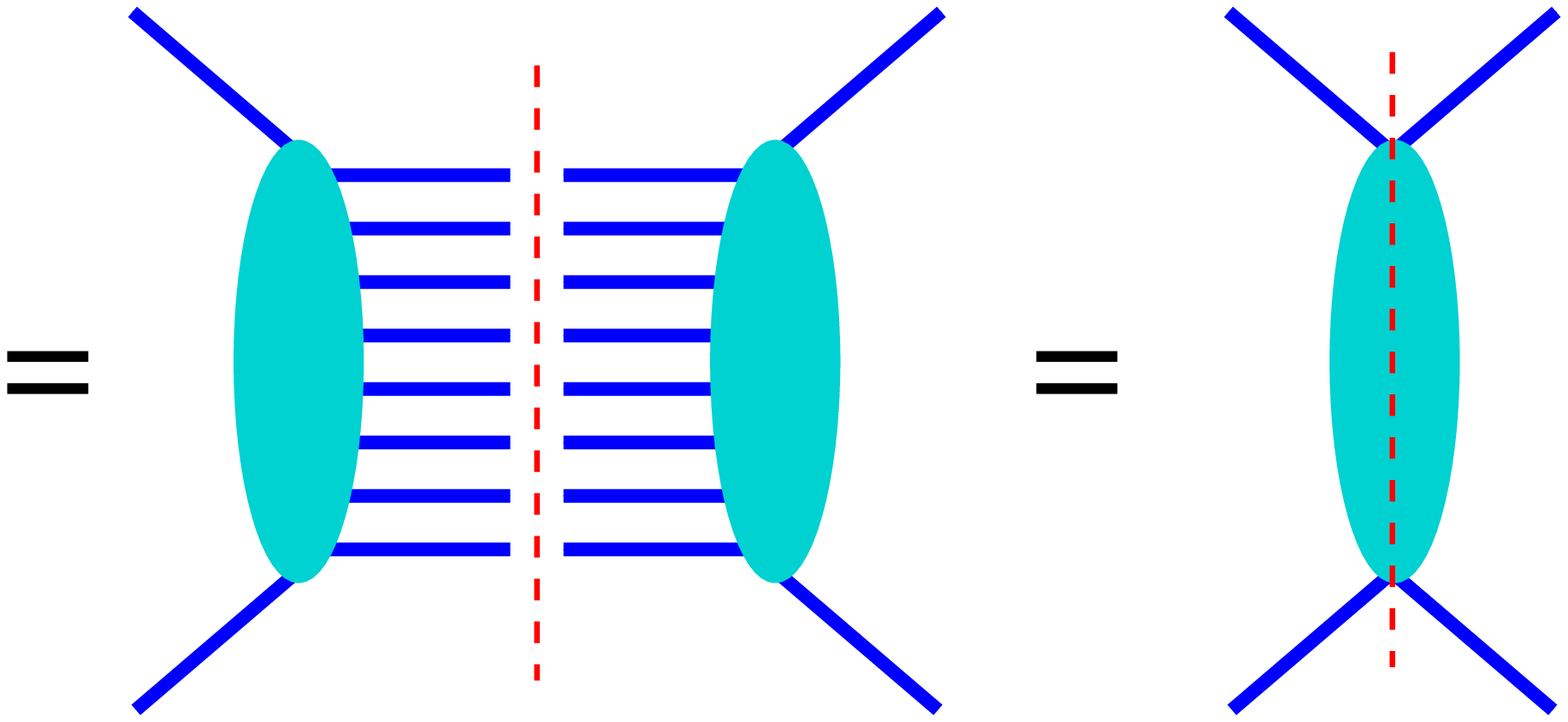}} \par}

\caption{The expression \protect\( \sum _{X}(T_{2\rightarrow X}).\protect \)\protect\( (T_{2\rightarrow X})^{*}\protect \)which
may be represented as a ``cut diagram''.\label{t2}}
\end{figure}

\begin{figure}[htb]
{\par\centering \resizebox*{!}{0.09\textheight}{\includegraphics{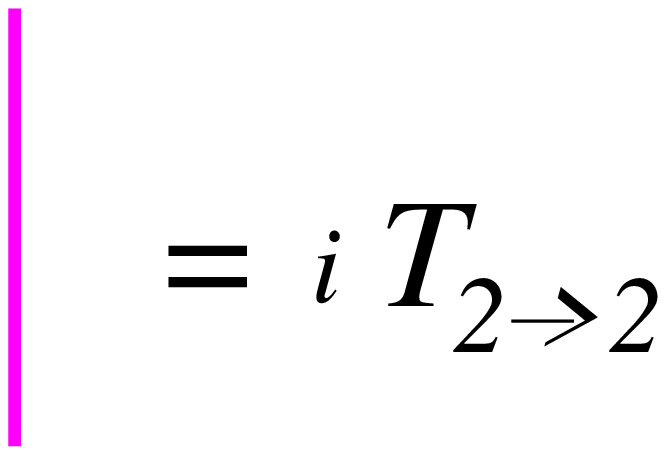}} \( \qquad  \)\resizebox*{!}{0.09\textheight}{\includegraphics{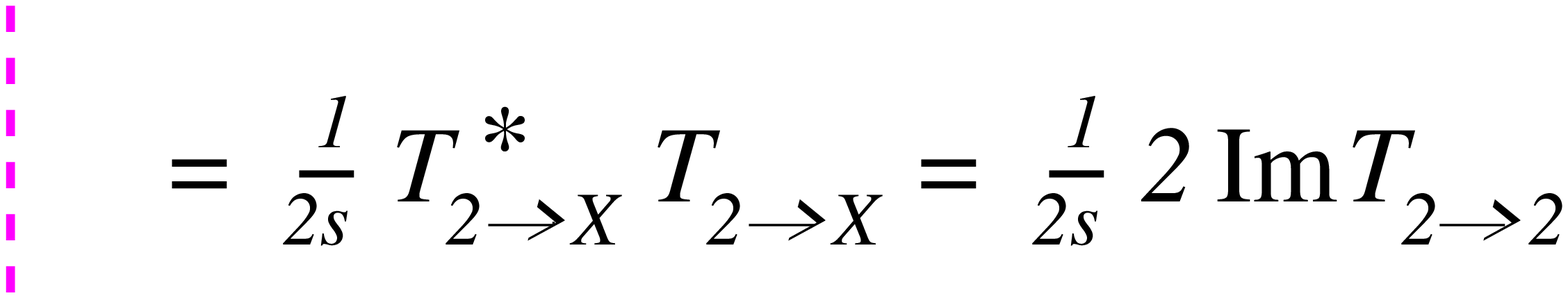}} \par}

\caption{Conventions..\label{line}}
\end{figure}

Let us introduce some conventions. We denote elastic two body scattering amplitudes
as \( T_{2\rightarrow 2} \) and inelastic amplitudes corresponding to the production
of some final state \( X \) as \( T_{2\rightarrow X} \) (see fig.\ref{t0}
). 
As a direct consequence of unitarity on has \( 2\, \mathrm{Im}T_{2\rightarrow 2}=\sum _{x} \)\( (T_{2\rightarrow X}) \)\( (T_{2\rightarrow X})^{*} \).
The right hand side of this equation may be literally presented as a ``cut
diagram'', where the diagram on one side of the cut is \( (T_{2\rightarrow X}) \)
and on the other side \( (T_{2\rightarrow X})^{*} \), as shown in fig.\ref{t2}
. 
So the term ``cut diagram'' means nothing but the square of an inelastic amplitude,
summed over all final states, which is equal to twice the imaginary part of
the elastic amplitude. Based on these considerations, we introduce simple graphical
symbols, which will be very convenient when discussing multiple scattering,
shown in fig. \ref{line}: a vertical solid line represents an elastic amplitude
(multiplied by \( i \), for convenience), and a vertical dashed line represents
the mathematical expression related to the above-mentioned cut diagram (divided
by \( 2s \), for convenience). 

\subsection{Elementary Interactions}

\begin{figure}[htb]
{\par\centering \resizebox*{!}{0.2\textheight}{\includegraphics{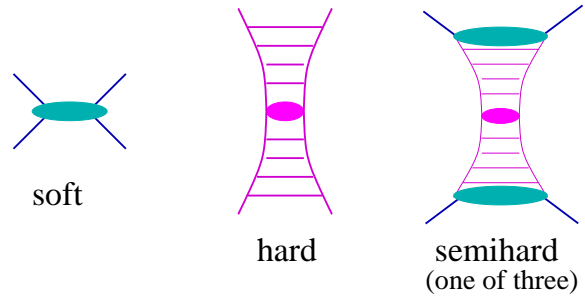}} \par}

\caption{The elastic amplitude \protect\( T_{2\rightarrow 2}\protect \).\label{t22}}
\end{figure}

Elementary nucleon-nucleon scattering can be considered as a straightforward
generalization of photon-nucleon scattering: one has a hard parton-parton scattering
in the middle, and parton evolutions in both directions towards the nucleons.
We have a hard contribution \( T_{\mathrm{hard}} \) when the the first partons
on both sides are valence quarks, a semi-hard contribution \( T_{\mathrm{semi}} \)
when at least on one side there is a sea quark (being emitted from a soft Pomeron),
and finally we have a soft contribution, when there is no hard scattering at
all (see fig. \ref{t22}). The total elementary elastic amplitude \( T_{2\rightarrow 2} \)
is the sum of all these terms.
We have a smooth transition from soft to hard physics: at low energies the soft
contribution dominates, at high energies the hard and semi-hard ones, at intermediate
energies (that is where experiments are performed presently) all contributions
are important. 

The multiple scattering theory will be based on these elementary interactions,
the corresponding elastic amplitude \( T_{2\rightarrow 2} \) and the corresponding
cut diagram, both being represented graphically by a solid and a dashed vertical
line. We also refer to the solid line as Pomeron, to the dashed line as cut
Pomeron.

\subsection{Multiple Scattering }

\begin{figure}[htb]
{\par\centering \resizebox*{!}{0.1\textheight}{\includegraphics{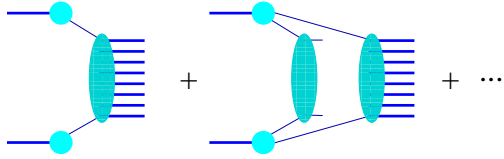}} \par}

\caption{Inelastic scattering in pp. \label{t7}}
\end{figure}

\begin{figure}[htb]
{\par\centering (a)\resizebox*{!}{0.1\textheight}{\includegraphics{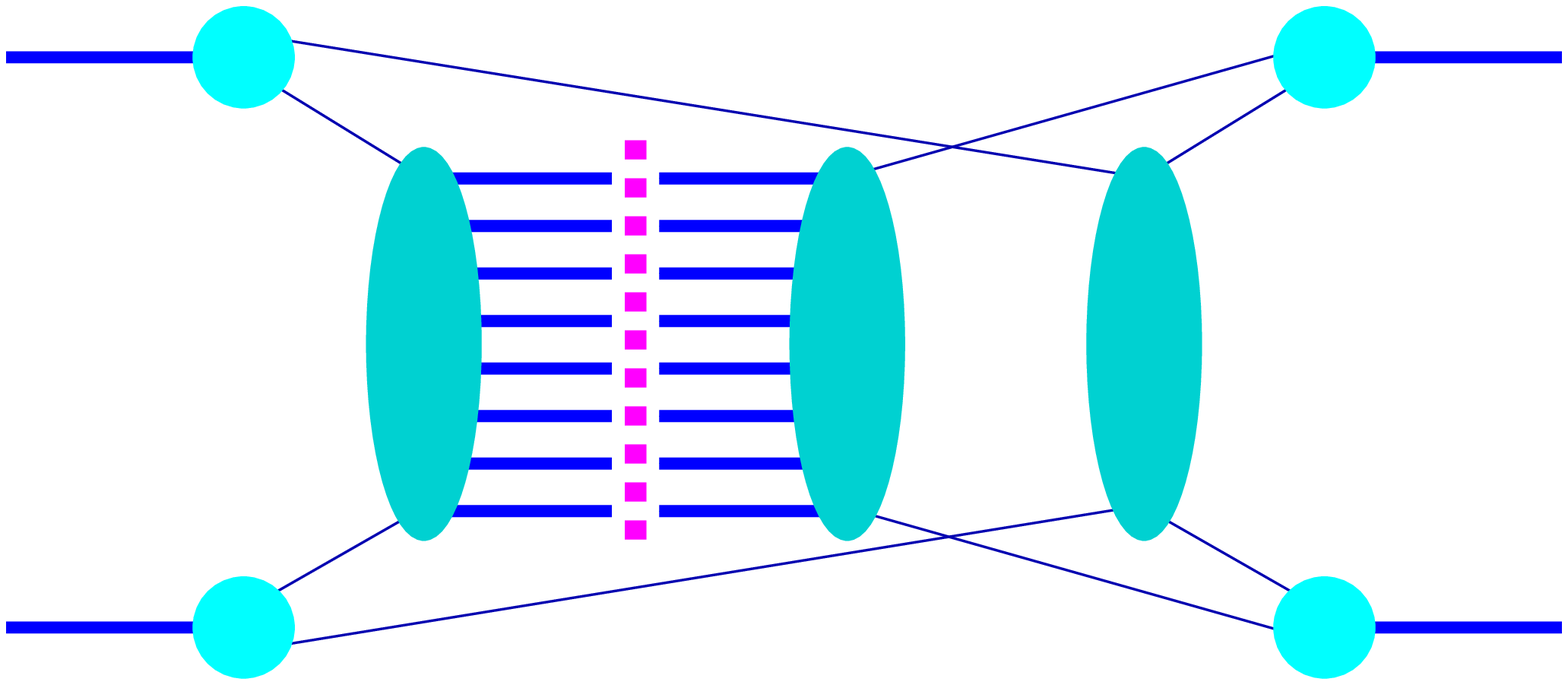}} \( \qquad  \)(b)\resizebox*{!}{0.1\textheight}{\includegraphics{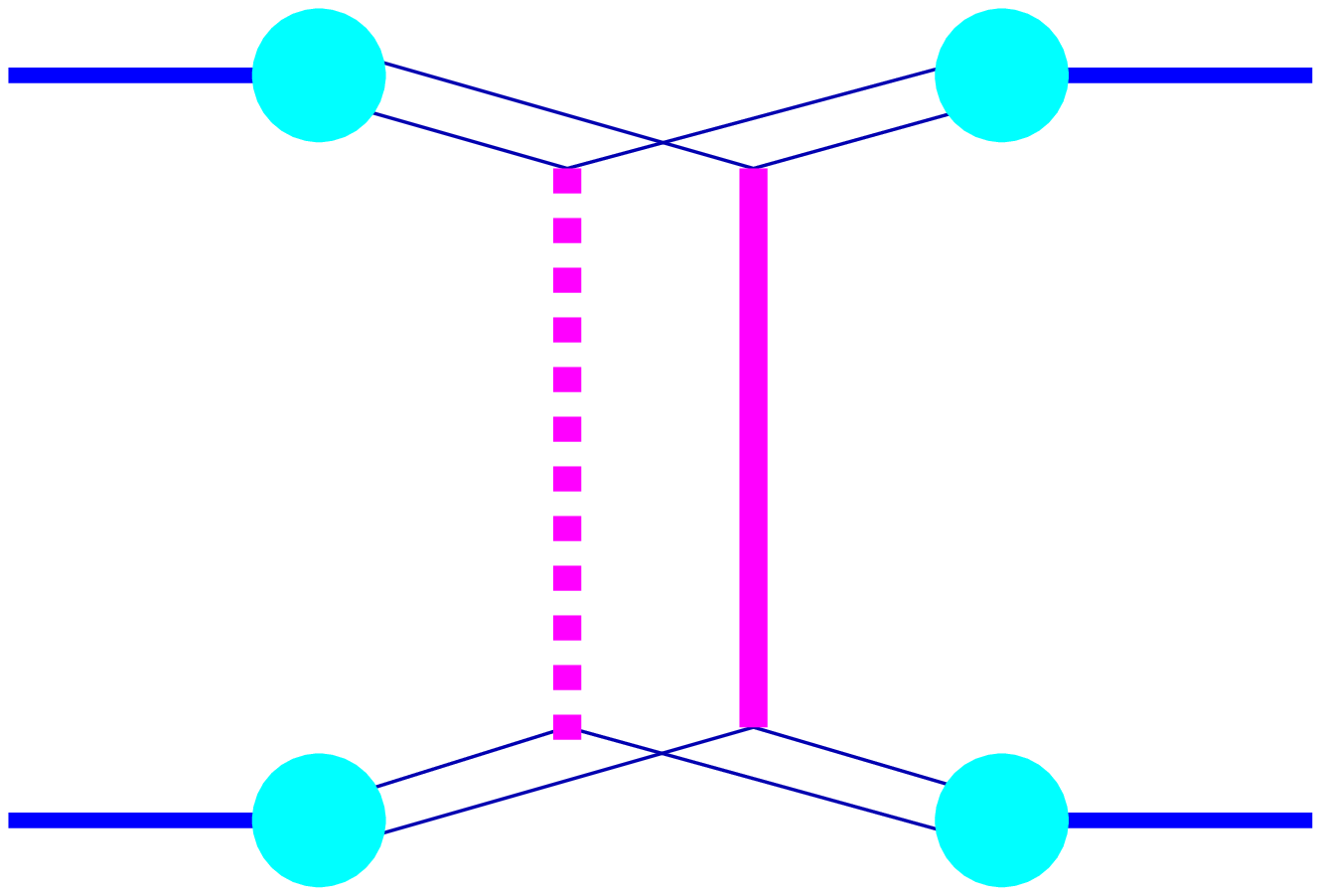}} \par}

\caption{Inelastic scattering in pp.  a) Amplitude, b) Squared amplitude makes\label{t7b}}
\end{figure}

We first consider inelastic proton-proton scattering, see fig. \ref{t7}. We
imagine an arbitrary number of elementary interactions to happen in parallel,
where an interaction may be elastic or inelastic. The inelastic amplitude is
the sum of all such contributions with at least one inelastic elementary interaction
involved. 
To calculate cross sections, we need to square the amplitude, which leads to
many interference terms, as the one shown in fig. \ref{t7b}(a), which represents
interference between the first and the second diagram of fig. \ref{t7}. Using
the above notations, we may represent the left part of the diagram as a cut
diagram, conveniently plotted as a dashed line, see fig. \ref{t7b}(b). The
amplitude squared is now the sum over many such terms represented by solid and
dashed lines.

\begin{figure}[htb]
{\par\centering \resizebox*{!}{0.08\textheight}{\includegraphics{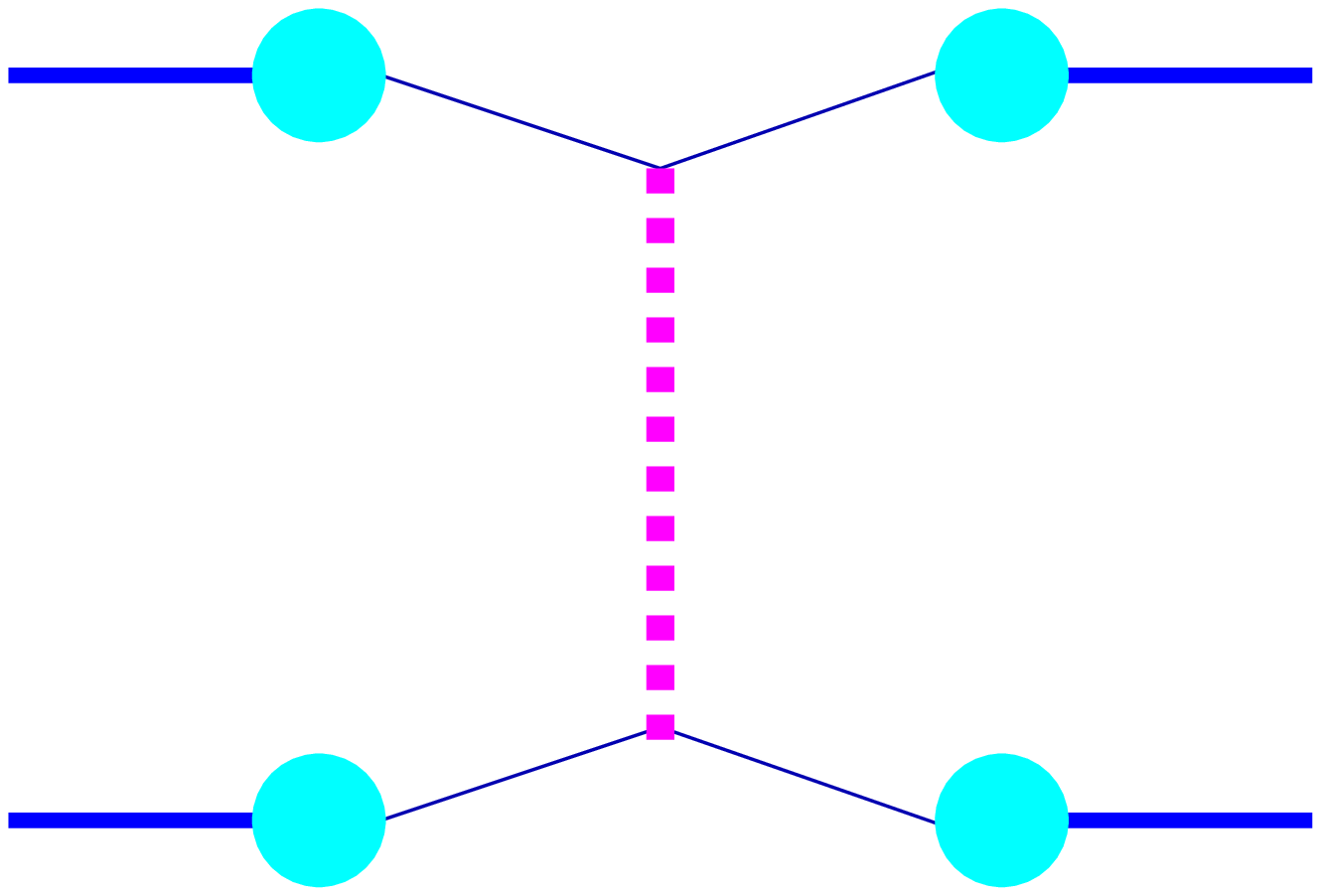}} \( \qquad  \)\resizebox*{!}{0.08\textheight}{\includegraphics{EPS/t7.eps}} \( \qquad  \)\resizebox*{!}{0.08\textheight}{\includegraphics{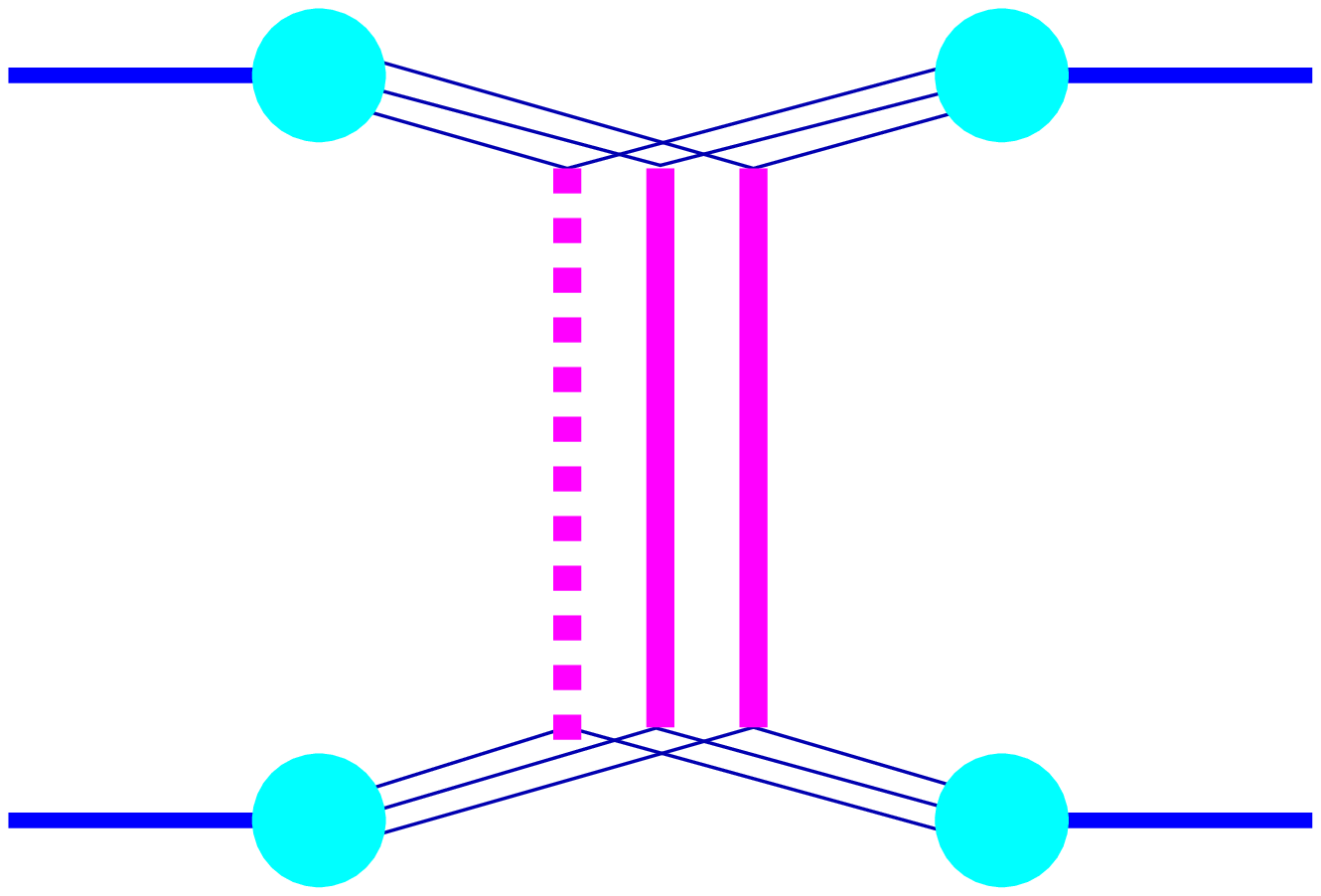}} \par}

\caption{Class of terms corresponding to one inelastic interaction.\label{t7c}}
\end{figure}

\begin{figure}[htb]
{\par\centering \resizebox*{!}{0.08\textheight}{\includegraphics{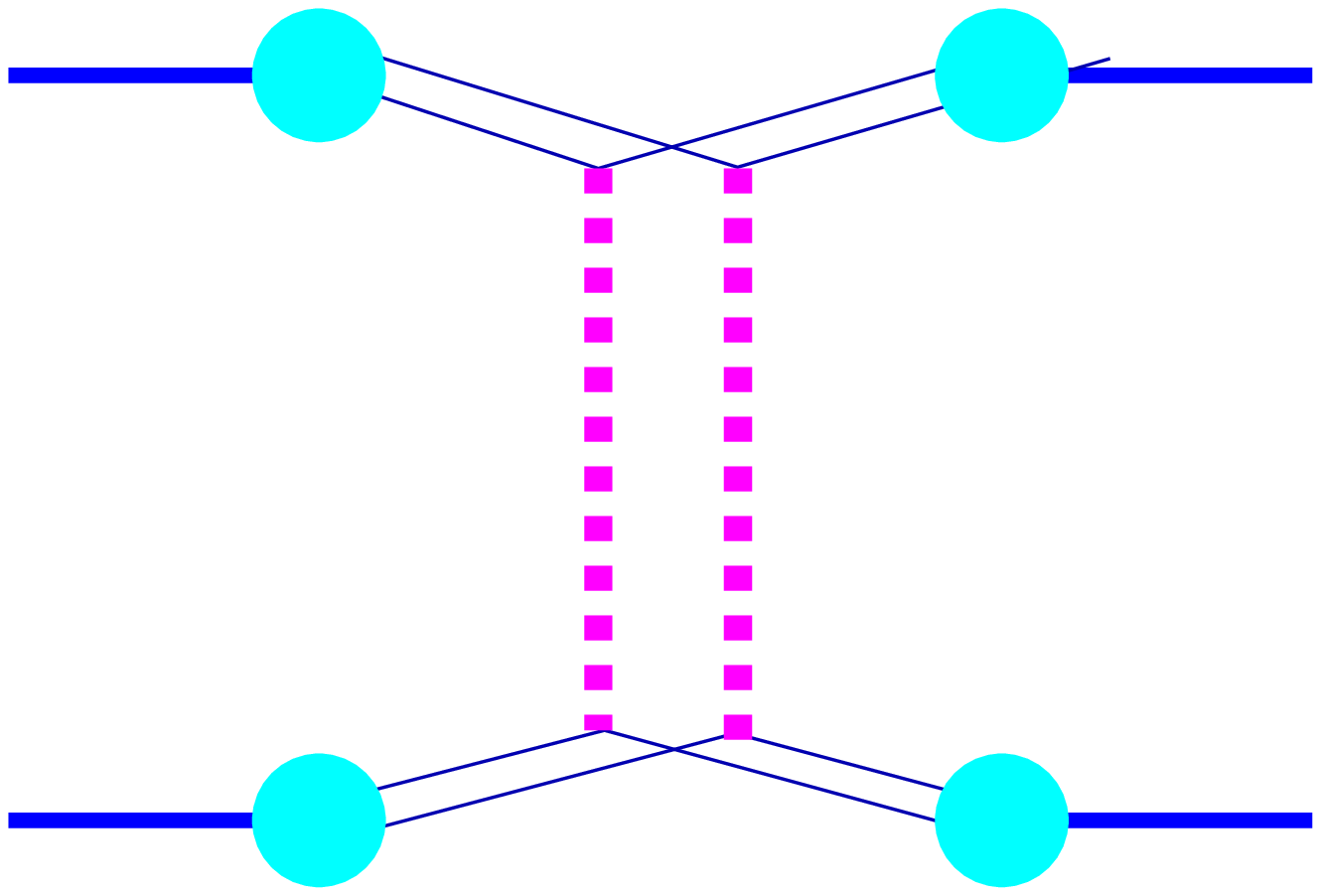}} \( \qquad  \)\resizebox*{!}{0.08\textheight}{\includegraphics{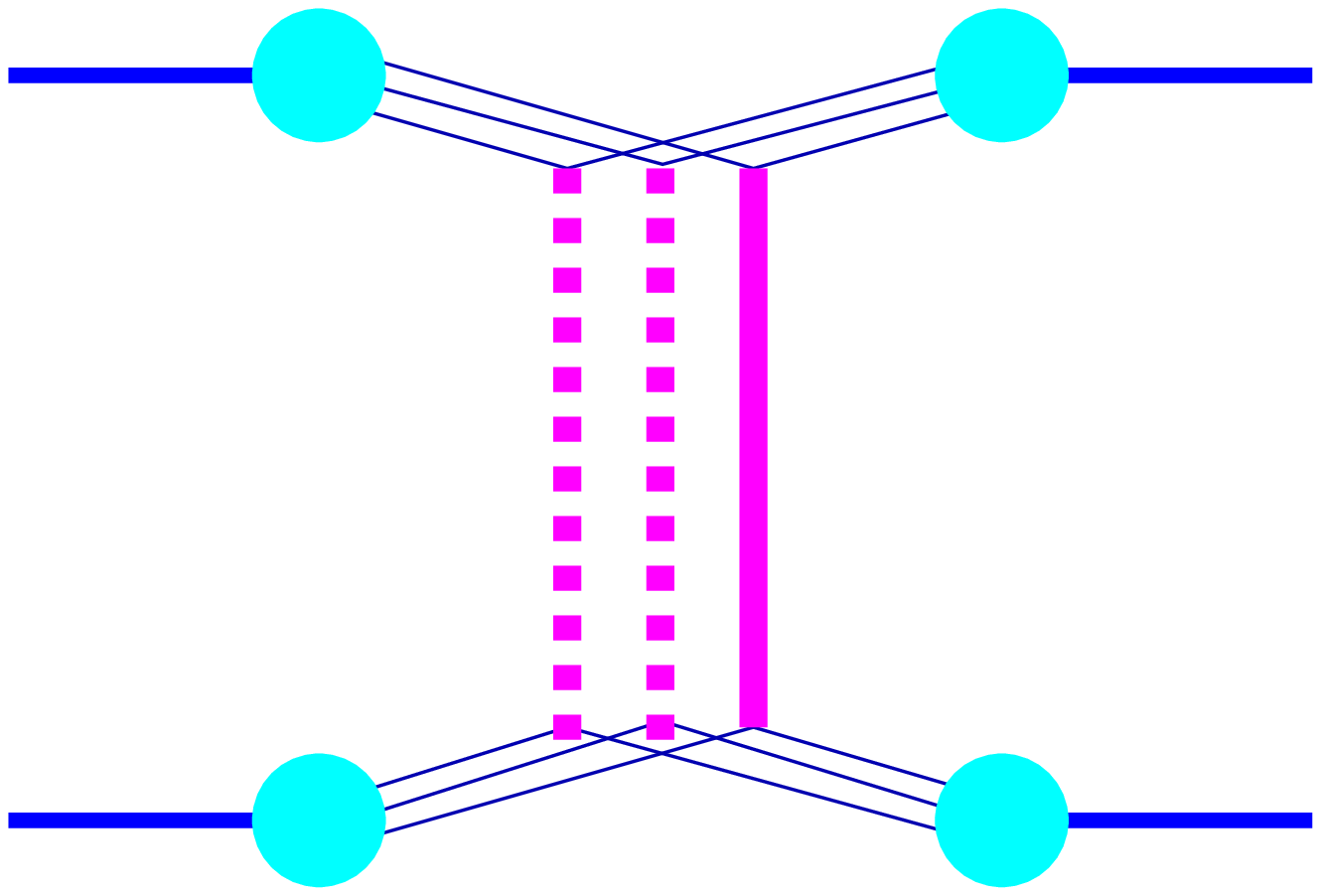}} \( \qquad  \)\resizebox*{!}{0.08\textheight}{\includegraphics{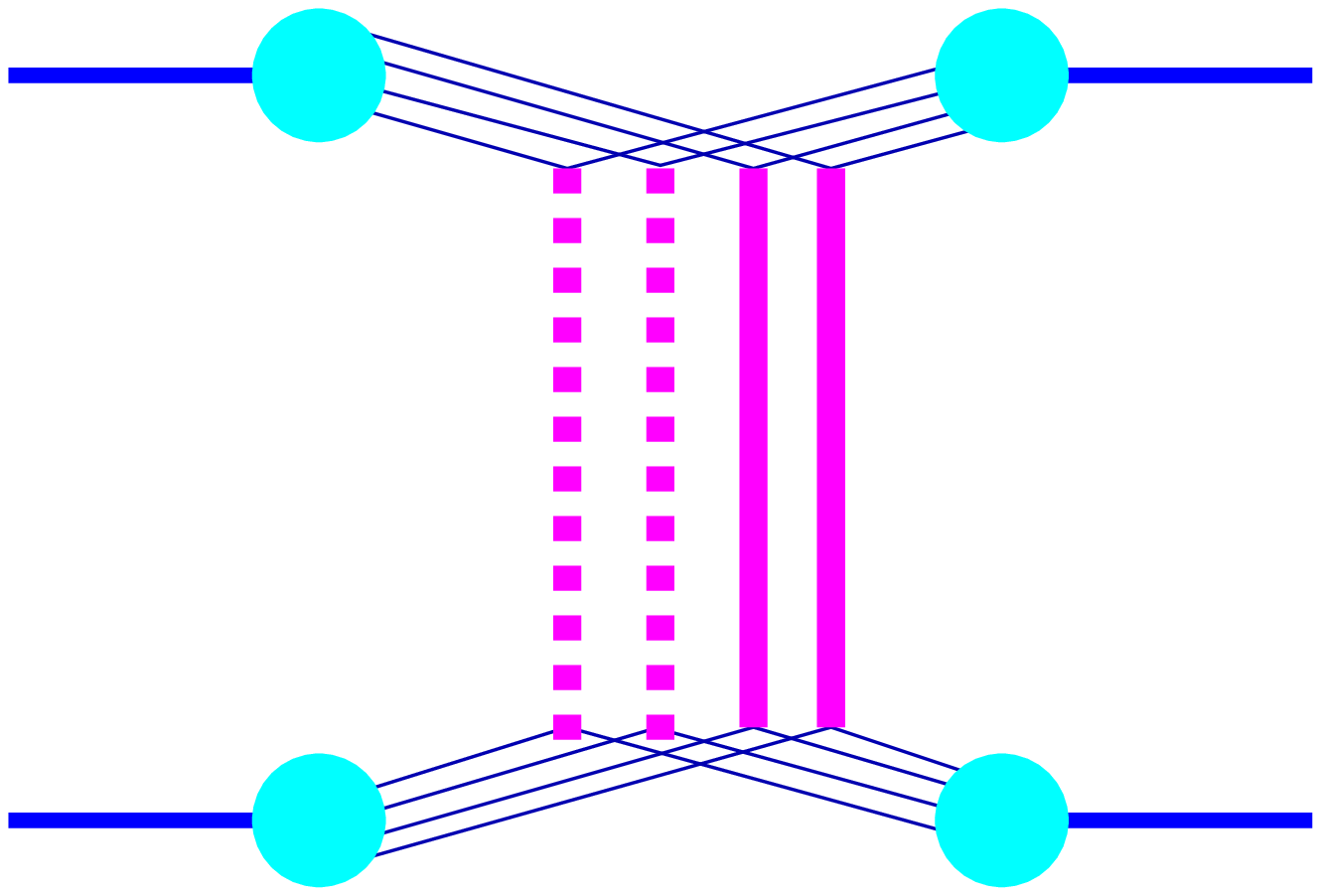}} \par}

\caption{Class of terms corresponding to two inelastic interactions.\label{t8c}}
\end{figure}

When squaring an amplitude being a sum of many terms, not all of the terms interfere
-- only those which correspond to the same final state. For example, a single
inelastic interaction does not interfere with a double inelastic interaction,
whereas all the contributions with exactly on inelastic interaction interfere.
So considering a squared amplitude, one may group terms together representing
the same final state. In our pictorial language, this means that all diagrams
with one dashed line, representing the same final state, may be considered to
form a class, characterized by \( m=1 \) -- one dashed line ( one cut Pomeron)
-- and the light cone momenta \( x^{+} \) and \( x^{-} \) attached to the
dashed line (defining energy and momentum of the Pomeron). In fig. \ref{t7c},
we show several diagrams belonging to this class, in fig. \ref{t8c}, we show
the diagrams belonging to the class of two inelastic interactions, characterized
by \( m=2 \) and four light-cone momenta \( x_{1}^{+} \), \( x_{1}^{-} \),
\( x_{2}^{+} \), \( x_{2}^{-} \).
Generalizing these considerations, we may group all contributions with \( m \)
inelastic interactions (\( m \) dashed lines = \( m \) cut Pomerons) into
a class characterized by the variable
\[
K=\{m,x_{1}^{+},x_{1}^{-},\cdots ,x_{m}^{+},x_{m}^{-}\}.\]
 We then sum all the terms in a class \( K \),
\[
\Omega (K)=\sum \{\mathrm{all}\, \mathrm{terms}\, \mathrm{in}\, \mathrm{class}\, K\}.\]
The cross section is then simply a sum over classes,
\[
\sigma _{\mathrm{inel}}(s)=\sum _{K\neq 0}\int d^{2}b\, \Omega (K).\]
\( \Omega  \) depends implicitly on the energy squared \( s \) and the impact
parameter \( b \). The individual terms \( \int d^{2}b\, \Omega (K) \), represent
partial cross sections, since they represent distinct final states. They are
referred to as topological cross sections. 

\begin{figure}[htb]
{\par\centering \resizebox*{!}{0.2\textheight}{\includegraphics{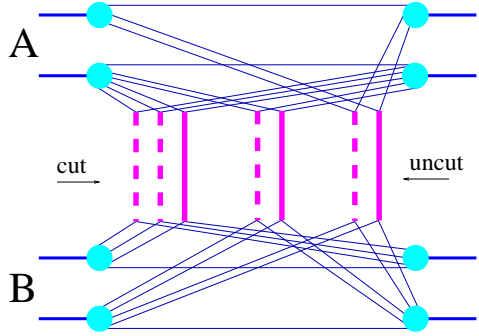}} \par}

\caption{An interference term of total cross section of \label{grtpabaa2}}
\end{figure}

The above concepts are easily generalized to nucleus-nucleus scattering, an
example for a diagram representing a contribution to the squared amplitude is
shown in fig. \ref{grtpabaa2}.
We may also define classes, which correspond to well defined final states, in
our notation a given number of dashed lines between nucleon pairs. We may number
the pairs as 1, 2, 3, ... \( k \) ... , \( AB \). We define \( m_{k} \) to
be the number of inelastic interactions (cut Pomerons) of the pair number \( k \).
The \( \mu ^{\mathrm{th}} \) of these \( m_{k} \) cut Pomerons is characterized
by light cone momenta \( x^{+}_{k\mu } \), \( x^{-}_{k\mu } \). So a class
may be characterized by
\[
K=\{m_{k},x^{+}_{k\mu },x^{-}_{k\mu }\}.\]
 We sum all terms in a class to obtain again a quantity called \( \Omega (K), \)
such that the cross section can be written as a sum over classes
\[
\sigma _{\mathrm{inel}}(s)=\sum _{K\neq 0}\int d^{2}b\, \Omega (K),\]
as in the case of proton-proton scattering. Here, however, \( b \) is a multidimensional
variable representing the impact parameter \( b_{0} \) and the transverse distances
\( b_{k} \) of all the nucleon-nucleon pairs. One can prove
\[
\sum _{K}\Omega (K)=1,\]
which is a very important result justifying our interpretation of \( \Omega (K) \)
to be a probability distribution for the configurations \( K \). This provides
also the basis for applying Monte Carlo techniques.

The function \( \Omega  \) is the basis of all applications of this formalism.
It provides the basis for calculating (topological) cross sections, but also
for particle production, thus providing a consistent formalism for all aspects
of a nuclear collision.

\subsection{Pomeron-Pomeron Interactions}

\begin{figure}[htb]
{\par\centering \resizebox*{!}{0.1\textheight}{\includegraphics{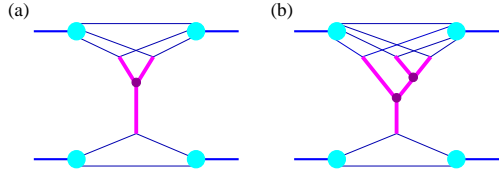}} \par}

\caption{Pomeron-Pomeron interactions.         \label{y}}
\end{figure}

\begin{figure}[htb]
{\par\centering \resizebox*{!}{0.12\textheight}{\includegraphics{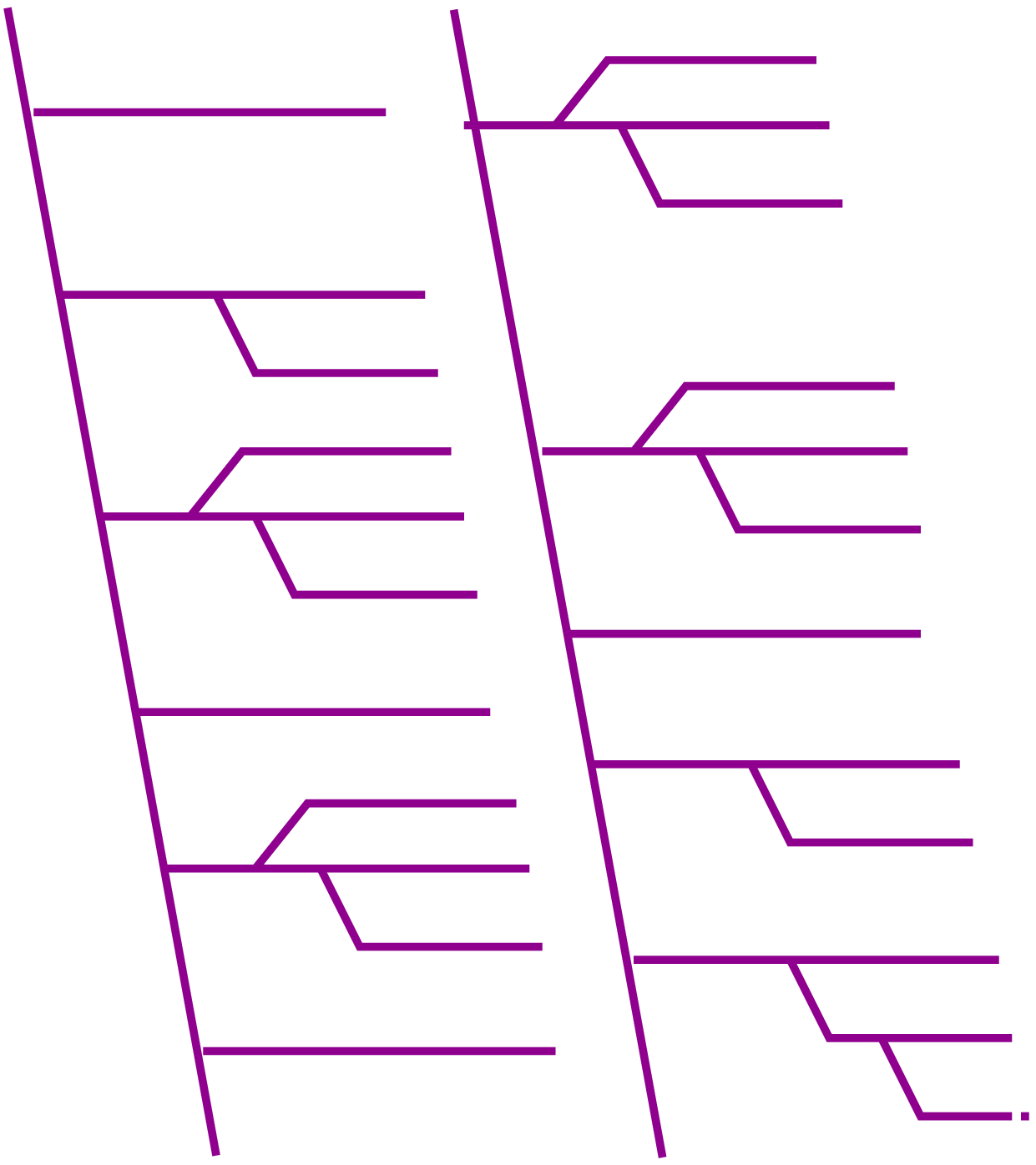}} \( \qquad  \)\resizebox*{!}{0.12\textheight}{\includegraphics{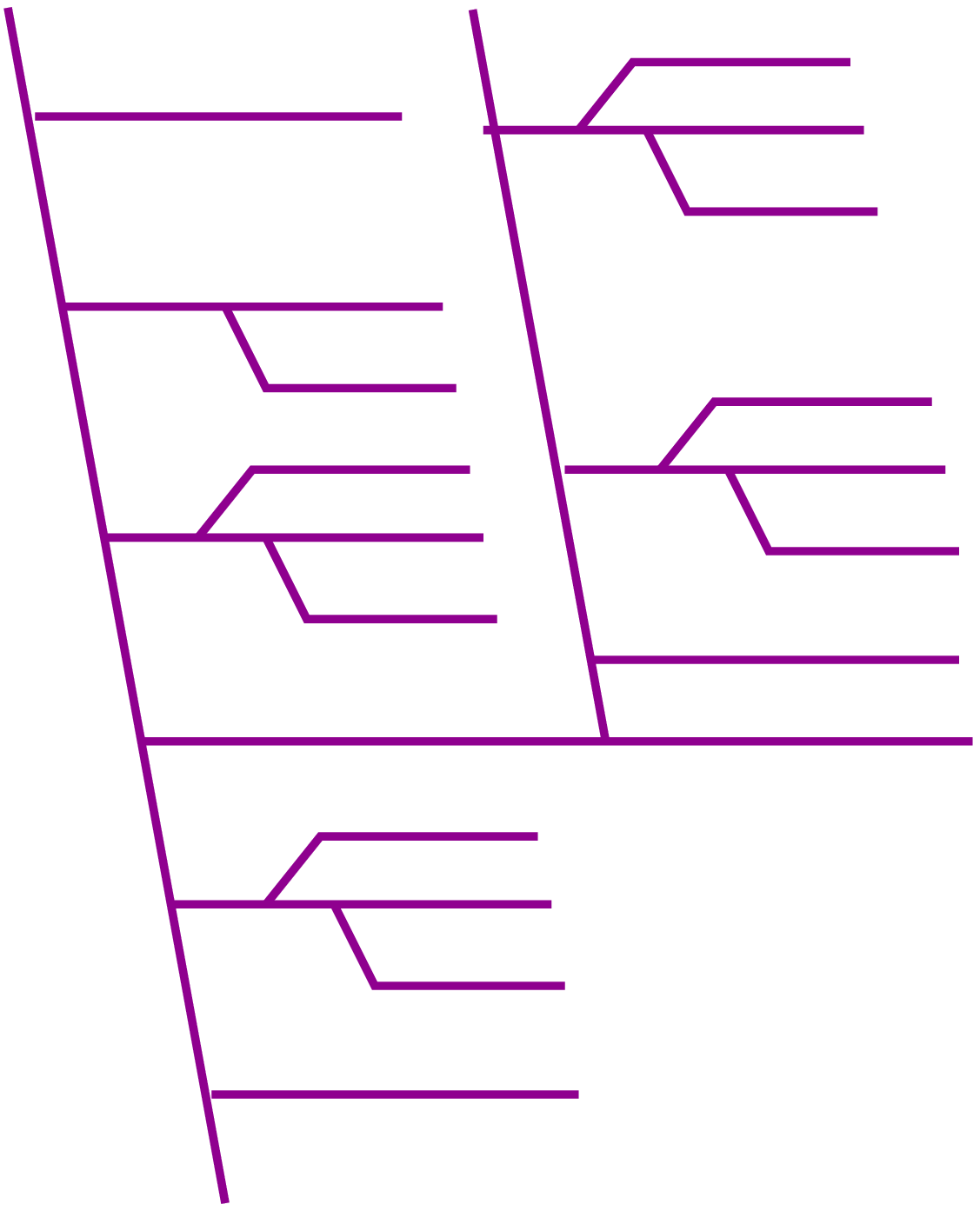}} \par}

\caption{Pomeron-Pomeron interactions in parton language. \label{cascade2}}
\end{figure}

So far, we consider the case where particle production from the individual elementary
interactions is completely independent. At high energies with high particle
densities this is not very realistic: particles emitted in one interaction could
be absorbed in another one. In our language: we have to allow interactions of
Pomerons, like the diagrams shown in fig. \ref{y}.
Such interactions are very important, being in particular responsible for screening
(shadowing, saturation). If we assume for a moment that a Pomeron is roughly
a parton ladder, then we we have the situation as shown in fig. \ref{cascade2}:
independent Pomerons correspond to non-interacting parton ladders (left figure),
whereas Pomeron interaction amount to interactions of partons from one ladder
with the ones from the other one (right figure). It is clear: the more partons
are produced, the more likely are such processes. 

\begin{figure}[htb]
{\par\centering \resizebox*{!}{0.1\textheight}{\includegraphics{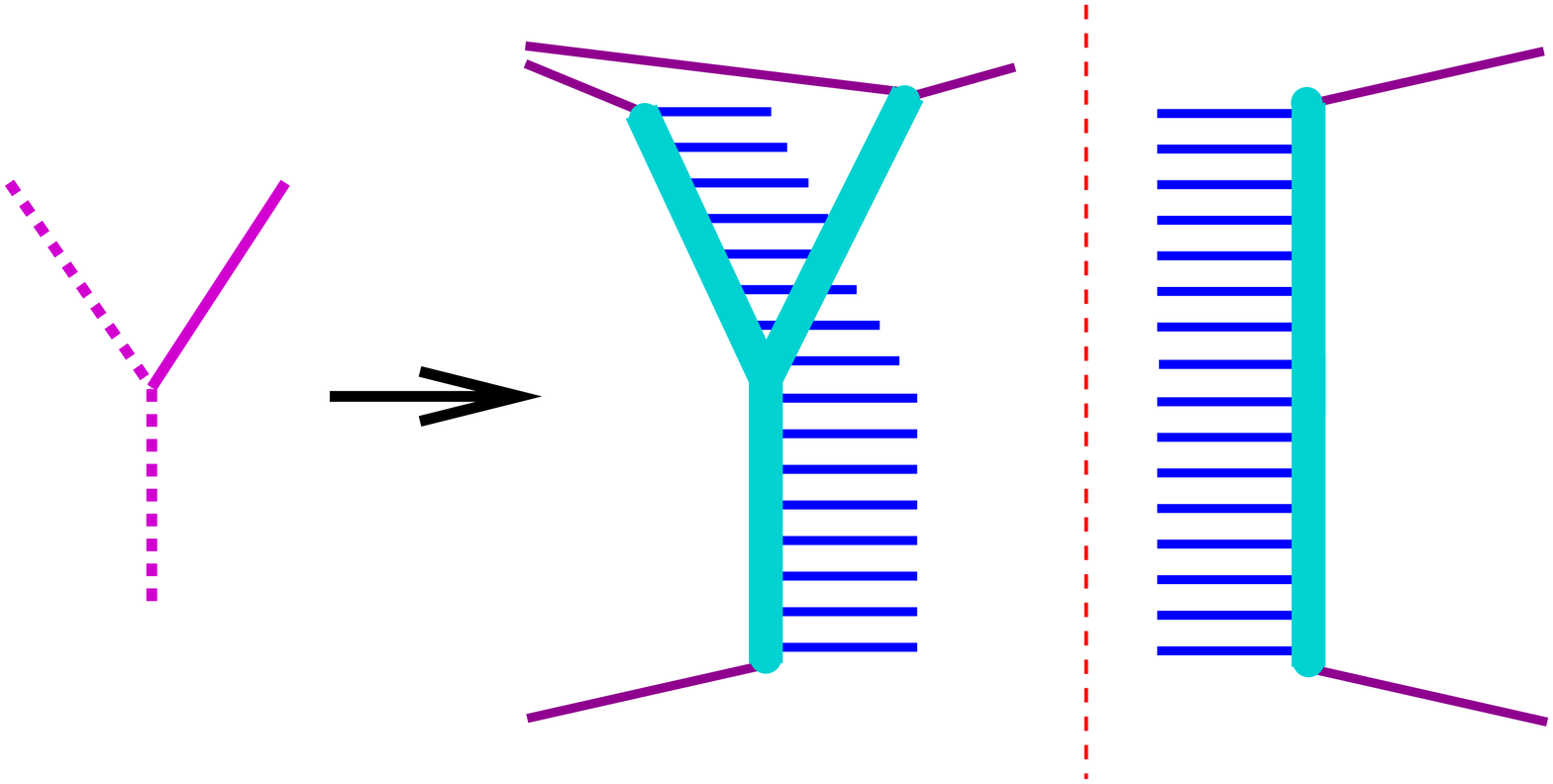}} \( \qquad  \)\resizebox*{!}{0.1\textheight}{\includegraphics{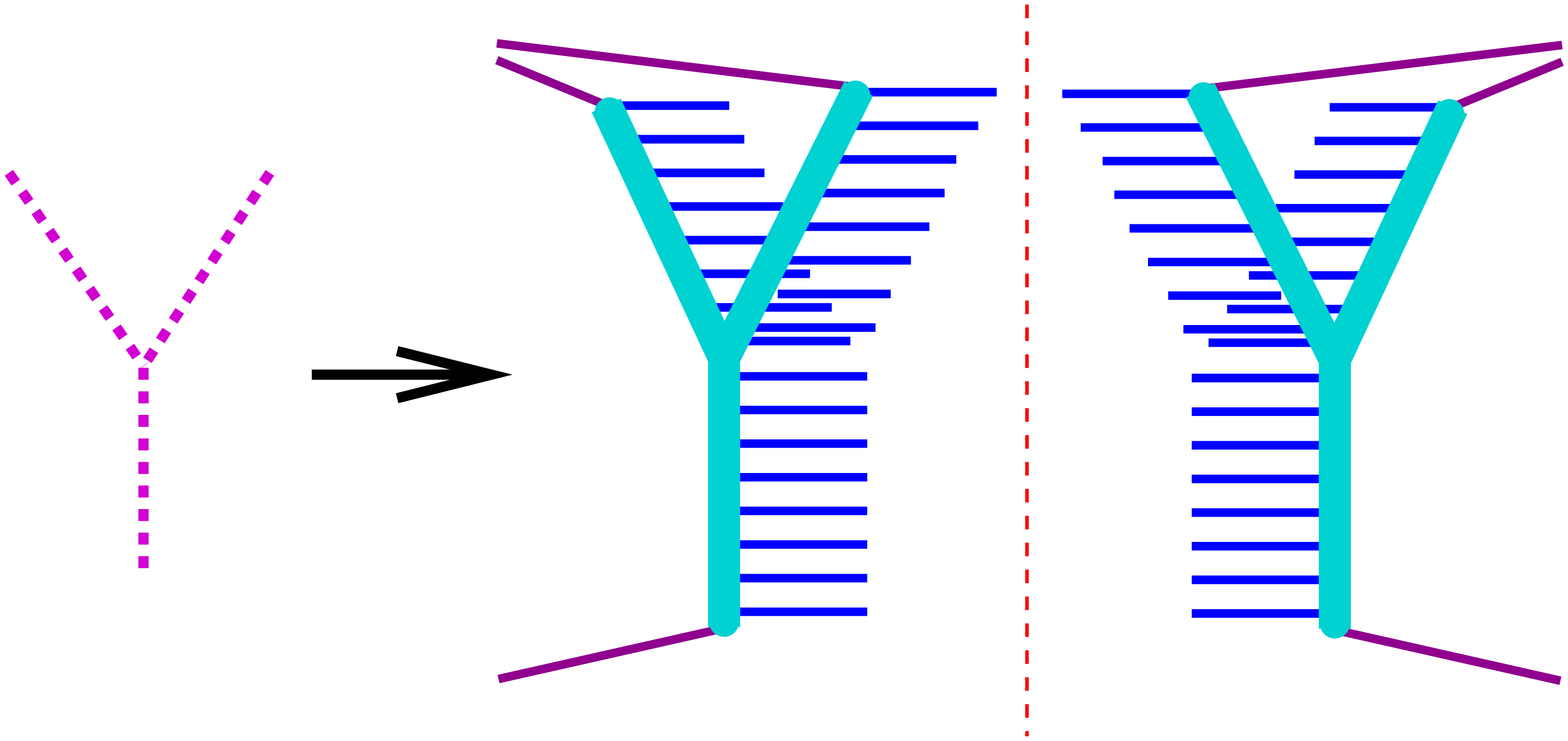}} \par}

\caption{Cut diagrams as a result of squared amplitudes..        \label{ty2}}
\end{figure}

\begin{figure}[htb]
{\par\centering \resizebox*{!}{0.12\textheight}{\includegraphics{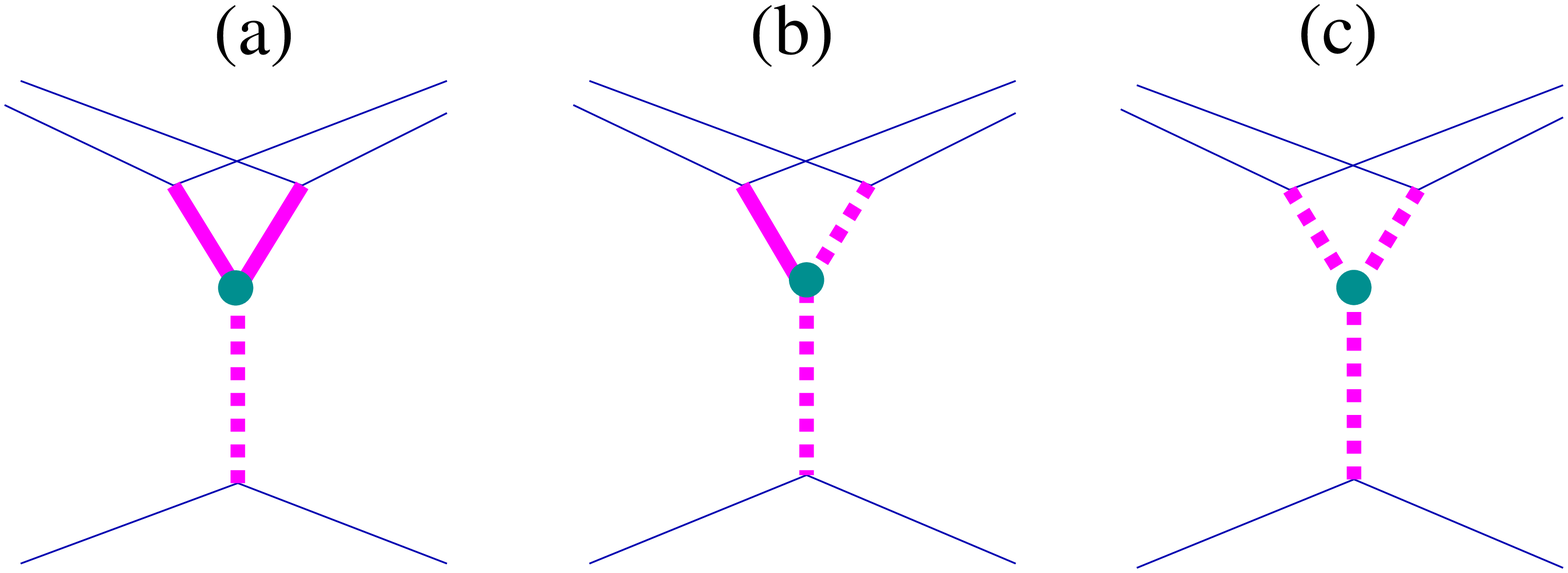}} \par}

\caption{Cut Y diagrams.         \label{ycut}}
\end{figure}

Also in case of Pomeron-Pomeron interactions, we are interested in particle
production, and so we have to worry about cutting diagrams. Again, cut diagrams
are the consequence of squaring amplitudes, i.e. multiplying an amplitude corresponding
to some process with the complex conjugate amplitude corresponding to the same
or some other process. In fig. \ref{ty2}, we show two examples: a ladder with
an additional leg is multiplied with a simple ladder (left figure), and two
ladders fused into one are multiplied with itself (right figure), We use again
dashed and solid lines for cut and uncut diagrams.
There are three cut diagrams of a Y diagram, as shown in fig. \ref{ycut}: the
lower leg is always cut; in addition, there may be none (a) or one (b) or two
(c) of the upper leg being cut.

\begin{figure}[htb]
{\par\centering \resizebox*{!}{0.25\textheight}{\includegraphics{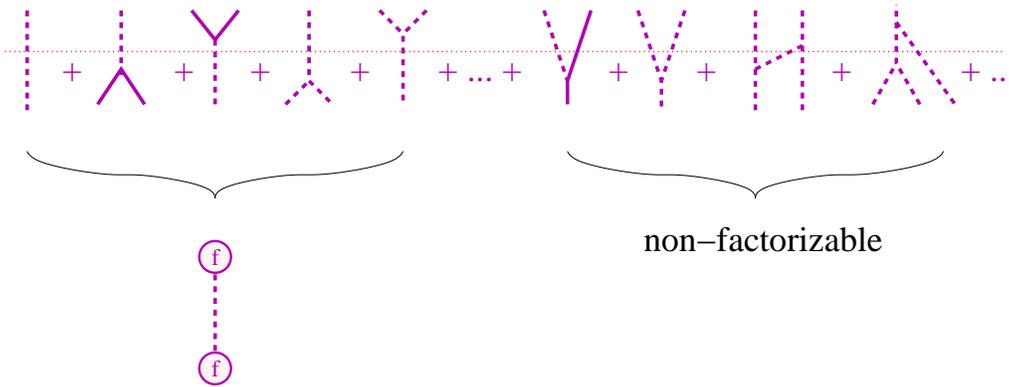}} \par}

\caption{Factorization in \protect\( pp\protect \) scattering.\label{yfactor5}}
\end{figure}

\begin{figure}[htb]
{\par\centering \ref{yfactor6}\par}

{\par\centering \resizebox*{!}{0.1\textheight}{\includegraphics{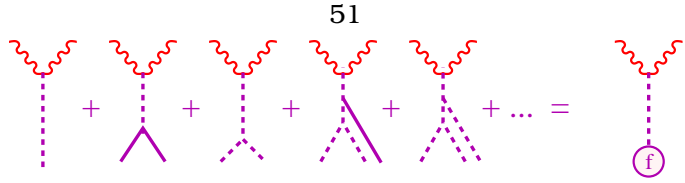}} \par}

\caption{Factorization in deep inelastic scattering. \label{yfactor6}}
\end{figure}

An important property of this formalism is the so-called factorization. A dashed
line corresponds to a cut Pomeron with given light cone momentum fractions \( x^{+} \)
and \( x^{-} \). So the rapidity of the Pomeron is \( 1/2\, \mathrm{ln}\, \mathrm{x}^{+}/x^{-} \),
the squared energy is \( s\, x^{+}\, x^{-} \). Suppose this cut Pomeron represents
a chain of particles with a typical transverse mass \( m \). The range of rapidity
is roughly given as \( y^{-}<y<y^{+} \), with \( y^{\pm }=\pm \ln (\sqrt{s}x^{\pm }/m) \)
. So we may assign a vertical rapidity scale and draw the dashed vertical lines
exactly between \( y^{+} \) and \( y^{-} \). In fig. \ref{yfactor5}, the
diagrams have been plotted this way. The horizontal dashed line represents some
given rapidity \( y \). Due to some general rules, only those diagrams contribute
to inclusive particle production at rapidity \( y \), where exactly one line
crosses the horizontal one. These are also the ones which factorize: they may
be considered as a single line between two ``blobs \( (f) \)'', each blob
being an infinite sum, providing thus a simple effective diagram. All the non-factorizable
diagrams do not contribute to the inclusive cross sections. 
In the same way, the structure function in deep inelastic scattering exhibits
factorization, as shown in fig. \ref{yfactor6}, with the same blob \( (f) \)
as in \( pp \) scattering. 
 This allows to write the inclusive cross section in \( pp \) as \( f\times \hat{\sigma }\times f \),
where \( \hat{\sigma } \) represents the dashed line, and \( f \) is obtained
from deep inelastic scattering. Essentially we recover here the parton model.

Does this mean that one can hide all these complicated multiple scattering features
in one simple measurable function \( f \)? The answer is yes if one is only
interested in calculating inclusive spectra. However, the situation is completely
different when it comes to the total cross section, where we have to consider
all diagrams. The above-mentioned cancellations concern only inclusive cross
sections. In addition, for Monte Carlo applications, we need to evaluate topological
cross sections, related to the probabilities to certain configurations (defined
by the numbers of cut Pomerons). Here again, no cancellations apply, we need
to consider all diagrams. 

In this sense, the so-called eikonal approach is very questionable, where total
and topological cross sections are calculated based on inclusive ones, neglecting
all the non-factorizable contributions.

\section*{Acknowledgments}

I would like to thank Fuming LIU and C. Javier SOLANO for helping to prepare
this manuscript.

\bibliographystyle{pr2}
\bibliography{a}

\end{document}